\documentclass[journal, hidelinks]{IEEEtran}

\pdfoutput=1
\IEEEoverridecommandlockouts

\usepackage{xr-hyper}
\usepackage{acronym}
\usepackage{amsmath}
\usepackage{amssymb}
\usepackage[english]{babel}
\usepackage{caption}
\usepackage{cite}
\usepackage[shortlabels,inline]{enumitem}
\usepackage{etoolbox}
\usepackage{graphicx}
\usepackage{hyperref}
\usepackage{bookmark}
\usepackage{ifthen}
\usepackage{cleveref}
\usepackage[utf8]{inputenc}
\usepackage{lipsum}
\usepackage{siunitx}
\usepackage[dvipsnames]{xcolor}
\usepackage[T1]{fontenc}
\usepackage[subnum]{cases}

\usepackage{epstopdf}
\graphicspath{{./fig/}}
\usepackage[toc,symbols]{glossaries}
\usepackage{bm}

\newtheorem{corollary}{Corollary}

\newtheorem{lem}{Lemma}

\newtheorem{prop}{Proposition}

\newcommand{\Figure}[1]{Fig.~\ref{#1}}
\newcommand{\Figures}[2]{Figs.~\ref{#1} and~\ref{#2}}
\newcommand{\FigureList}[2]{Figs.~\ref{#1} --~\ref{#2}}

\newcommand{\Equation}[1]{\eqref{#1}}
\newcommand{\Equations}[2]{\eqref{#1} and~\eqref{#2}}
\newcommand{\EquationsList}[2]{\eqref{#1} --~\eqref{#2}}
\newcommand{\Table}[1]{Table~\ref{#1}}

\newcommand{\Section}[1]{Section~\ref{#1}}

\newcommand{\Appendix}[1]{Appendix~\ref{#1}}

\newcommand{\Proposition}[1]{Proposition~\ref{#1}}
\newcommand{\Corollary}[1]{Corollary~\ref{#1}}

\newcommand\given[1][]{\:#1\vert\:}
\newcommand\prob[1]{\textnormal{Pr}\{{#1}\}}
\newcommand\fpdf[2][]{\textnormal{f}_{\mathrm{#1}}\left(#2\right)}
\newcommand{\poisDist}[2]{\ensuremath{\frac{{(#1)}^{#2} \exp\left(-(#1)\right)}{#2!}}}
\newcommand{\pois}[1]{{\ensuremath{\textnormal{Pois}\left({#1}\right)}}}

\acrodef{2D}[2-D]{two-dimensional}
\acrodef{3D}[3-D]{three-dimensional}
\acrodef{MC}{molecular communication}
\acrodef{OOK}{ON-OFF keying}
\acrodef{ISI}{inter-symbol interference}
\acrodef{ILI}{inter-link interference}
\acrodef{IUI}{inter-user interference}
\acrodef{IR}{impulse response}
\acrodef{ML}{maximum likelihood}
\acrodef{wlog}[w.l.o.g.]{without loss of generality}
\acrodef{BER}{bit error rate}
\acrodef{SER}{symbol error rate}
\acrodef{BSC}{Binary Symmetric Channel}
\acrodef{CDF}{cumulative density function}
\acrodef{UCA}{uniform concentration assumption}
\acrodef{AWGN}{Additive White Gaussian Noise}
\acrodef{PBS}{particle-based simulation}
\acrodef{MLSE}{maximum likelihood sequence estimator}
\acrodef{VE}{viterbi equalizer}
\acrodef{MLE}{maximum likelihood estimator}
\acrodef{CIR}{channel impulse response}
\acrodef{CIRs}{channel impulse responses}
\acrodef{SNR}{signal-to-noise ratio}
\acrodef{SDMA}{Space Division Multiple Access}
\acrodef{MCDMA}{Molecular Code Division Multiple Access}
\acrodef{ADMA}{Amplitude-Division Multiple Access}
\acrodef{MTDMA}{Molecular Time Division Multiple Access}
\acrodef{MDMA}{Molecular Division Multiple Access}
\acrodef{MIMO}{multiple-input multiple-output}
\acrodef{TDMA}{Time Division Multiple Access}
\acrodef{CDMA}{Code Division Multiple Access}
\acrodef{CSI}{channel state information}
\acrodef{TX}{transmitter}
\acrodef{TXs}{transmitters}
\acrodef{RX}{receiver}
\acrodef{RXs}{receivers}
\acrodef{ARE}{area rate efficiency}
\acrodef{ARTE}{area and time rate efficiency}
\acrodef{w.r.t.}{with respect to}

\DeclareMathOperator\erf{erf}

\makeatletter
\long\def\@makecaption#1#2{\ifx\@captype\@IEEEtablestring%
    \footnotesize\begin{center}{\normalfont\footnotesize #1}\\
        {\normalfont\footnotesize\scshape #2}\end{center}%
    \@IEEEtablecaptionsepspace
    \else
    \@IEEEfigurecaptionsepspace
    \setbox\@tempboxa\hbox{\normalfont\footnotesize {#1.}~~ #2}%
    \ifdim \wd\@tempboxa >\hsize%
    \setbox\@tempboxa\hbox{\normalfont\footnotesize {#1.}~~ }%
    \parbox[t]{\hsize}{\normalfont\footnotesize \noindent\unhbox\@tempboxa#2}%
    \else
    \hbox to\hsize{\normalfont\footnotesize\hfil\box\@tempboxa\hfil}\fi\fi}
\makeatother

\newcommand{\scaleSection}{\vspace{-0cm}}
\newcommand{\scaleSubsection}{\vspace{-0cm}}
\newcommand{\scaleSubsubsection}{\vspace{-0cm}}
\newcommand{\scaleSectionBelow}{\vspace{-0cm}}
\newcommand{\scaleSubsectionBelow}{\vspace{-0cm}}
\newcommand{\scaleSubsubsectionBelow}{\vspace{-0cm}}
\newcommand{\scaleAlign}{\vspace{-0.1cm}}

\newglossaryentry{ringNr}{%
  name = {\ensuremath{N_{\textnormal{rings}}}},
  description = {the current maximum ring number of interferers},
  type= symbols
}

\newglossaryentry{r0minHex}{%
  name = {\ensuremath{c}},
  description = {the distance between two adjacent transmitters in a hexagonal grid},
  type= symbols
}

\newglossaryentry{nrMol}{%
  name={\ensuremath{N_{\textnormal{m}}}},
  description={ the number of released molecules at the transmitter, which is equal for all the transmitters},
  type=symbols
}

\newglossaryentry{SRX}{%
  name={\ensuremath{S_{\mathrm{RX}}}},
  description={ the radius of a receiver},
  type=symbols
}
\newglossaryentry{SRX2}{%
  name={\ensuremath{S^2_{\mathrm{RX}}}},
  description={ squared radius of a receiver},
  type=symbols
}

\newglossaryentry{LRX}{%
  name={\ensuremath{L_{\mathrm{RX}}}},
  description={ the length of a receiver},
  type=symbols
}

\newglossaryentry{VRX}{%
  name={\ensuremath{V_{\mathrm{RX}}}},
  description={ the volume of a receiver, $\gls{VRX} = \gls{LRX} \pi \gls{SRX}^2$},
  type=symbols
}

\newglossaryentry{flow}{%
  name={\ensuremath{v_{z}}},
  description={ flow that points into the $z$ direction, $\gls{flow} = \gls{flow}_z$},
  type=symbols
}

\newglossaryentry{diffusion}{%
  name={\ensuremath{D}},
  description={ diffusion coefficient},
  type=symbols
}

\newglossaryentry{d}{%
  name={\ensuremath{d}},
  description={ the distance between transmitter and receiver plane},
  type=symbols
}

\newglossaryentry{nrIntU}{%
  name={\ensuremath{N_{\textnormal{IU}}}},
  description= the number of interfering users,
  type=symbols
}

\newglossaryentry{nrIntUG}{%
  name={\ensuremath{N_{\textnormal{IG}}}},
  description= the number of interfering groups,
  type=symbols
}

\newglossaryentry{nrTrans}{%
  name={\ensuremath{N_{\textnormal{TX}}}},
  description= the number of transmitters,
  type=symbols
}

\newglossaryentry{curTrans}{%
  name={\ensuremath{i}},
  description= {current transmitter with $\gls{curTrans} \in [1, \gls{nrTrans}]$},
  type=symbols
}

\newglossaryentry{nrRec}{%
  name={\ensuremath{N_\textnormal{RX}}},
  description= the number of transmitters,
  type=symbols
}

\newglossaryentry{curRec}{%
  name={\ensuremath{j}},
  description= {current receiver with $\gls{curRec} \in [1, \gls{nrRec}]$},
  type=symbols
}

\newglossaryentry{RX0}{%
  name={\ensuremath{\acs{RX}_0}},
  description= {current center receiver that will be the example receiver throughout the paper},
  type=symbols
}

\newglossaryentry{TX0}{%
  name={\ensuremath{\acs{TX}_0}},
  description= {current center transmitter that will be the example partner transmitter to \gls{RX0} throughout the paper},
  type=symbols
}

\newglossaryentry{samplingTime}{%
  name={\ensuremath{t_{\textnormal{s}}}},
  description= the time at which a receiver counts the currently receiving molecules,
  type=symbols
}

\newglossaryentry{CIR}{%
  name={\ensuremath{{\textnormal{CIR}_{\gls{curTrans},\gls{curRec}}(t)}}},
  description= number of molecules from transmitter \gls{curTrans} at receiver \gls{curRec} over time,
  type=symbols
}

\newglossaryentry{csmean}{%
  name={\ensuremath{\overline{c}_{\mathrm{s}}}},
  description= {the expected number of molecules from the TX partner at sampling time $t = \gls{samplingTime}$ with $\gls{csmean} = \gls{nrMol} \gls{CIR}$ and $\gls{curTrans} = \gls{curRec}$},
  type=symbols
}

\newglossaryentry{cmeaniui}{%
  name={\ensuremath{\overline{\boldsymbol{c}}_{\textnormal{ILI}}}},
  description= {From the perspective of one receiver \gls{curRec} the expected number of molecules of all transmitters that are not the transmitter partner, $\gls{curTrans} \neq \gls{curRec}$ at sampling time $t = \gls{samplingTime}$ with $\gls{cmeaniui} = [\overline{c}_{1,\gls{curRec}}, \overline{c}_{2,\gls{curRec}}, \dots, \overline{c}_{\gls{curRec}-1,\gls{curRec}}, \overline{c}_{\gls{curRec}+1,\gls{curRec}}, \dots, \overline{c}_{\gls{nrTrans},\gls{curRec}}]$},
  type=symbols
}

\newglossaryentry{cmeaniuiTrans}{%
  name={\ensuremath{\overline{\boldsymbol{c}}^{\textnormal{T}}_{\textnormal{ILI}}}},
  description= {From the perspective of one receiver \gls{curRec} the expected number of molecules of all transmitters that are not the transmitter partner, $\gls{curTrans} \neq \gls{curRec}$ at sampling time $t = \gls{samplingTime}$ with $\gls{cmeaniui} = [\overline{c}_{1,\gls{curRec}}, \overline{c}_{2,\gls{curRec}}, \dots, \overline{c}_{\gls{curRec}-1,\gls{curRec}}, \overline{c}_{\gls{curRec}+1,\gls{curRec}}, \dots, \overline{c}_{\gls{nrTrans},\gls{curRec}}]$},
  type=symbols
}

\newglossaryentry{cmeann}{%
  name={\ensuremath{\overline{c}_{\mathrm{n}}}},
  description= {the expected number of background noise molecules at sampling time $t = \gls{samplingTime}$},
  type=symbols
}

\newglossaryentry{cmeanij}{%
  name={\ensuremath{\overline{c}_{\gls{curTrans},\gls{curRec}}}},
  description= {the expected number of molecules from transmitter \gls{curTrans} at receiver \gls{curRec} at sampling time $t = \gls{samplingTime}$ with $\gls{cmeanij} = \gls{nrMol} \gls{CIR}$},
  type=symbols
}

\newglossaryentry{cmeaniuiScalar}{%
name={\ensuremath{\overline{c}_{\textnormal{ILI}}}},
description= {From the perspective of one receiver \gls{curRec} the expectation over the expected number of molecules of all transmitters that are not the transmitter partner, $\gls{cmeaniuiScalar} = \underset{\gls{siui}}{\mathbb{E}}\{\gls{siui}\gls{cmeaniui}^{\textnormal{T}}\}$},
type=symbols
}

\newglossaryentry{ss}{%
  name={\ensuremath{s_0}},
  description={From the perspective of one receiver \gls{curRec}, the current symbol of the TX partner, $\gls{curTrans} = \gls{curRec}$, with $\gls{ss} = \{0,1\}$, so there are two states for \gls{ss}},
  type=symbols
}
\newglossaryentry{ssHat}{%
  name={\ensuremath{\hat{s}_0}},
  description={From the perspective of one receiver \gls{curRec}, the current symbol of the TX partner, $\gls{curTrans} = \gls{curRec}$, with $\gls{ss} = \{0,1\}$, so there are two states for \gls{ss}},
  type=symbols
}

\newglossaryentry{Riui}{%
  name={\ensuremath{N_{\textnormal{ILI}}}},
  description={number of possible states of the interference coming from other users with  $\gls{Riui} = 2^{\gls{nrTrans}-1}$},
  type=symbols
}

\newglossaryentry{siui}{%
  name={\ensuremath{\boldsymbol{s}_{\textnormal{ILI}}}},
  description={From the perspective of one receiver \gls{curRec}, the current symbols of all transmitters that are not the transmitter partner, $\gls{curTrans} \neq \gls{curRec}$, with$\gls{siui} = [s_1, s_2, \dots, s_{\gls{curRec}-1}, s_{\gls{curRec}+1}, \dots, s_{\gls{nrTrans}}]$ and $s_i = \{0,1\}$, so there are $\gls{Riui} = 2^{\gls{nrTrans}-1}$ realisations possible},
  type=symbols
}

\newglossaryentry{cs}{%
  name={\ensuremath{c_{\mathrm{s}}}},
  description= {a random number which represents the number of molecules from the TX partner at one receiver, $\gls{curTrans} = \gls{curRec}$ at sampling time \gls{samplingTime} with $\gls{cs} \sim \pois{\gls{ss}\gls{csmean} } = \fpdf{\gls{cs} \given \gls{ss} }$},
  type=symbols
}

\newglossaryentry{ciui}{%
  name={\ensuremath{c_{\textnormal{ILI}}}},
  description= {a random number which represents the number of molecules from the all transmitters that are not the transmitter partner, $\gls{curTrans} \neq \gls{curRec}$, at sampling time \gls{samplingTime} with $\gls{ciui} \sim \pois{\gls{siui}\gls{cmeaniui}^{\textnormal{T}} } = \fpdf{\gls{ciui} \given \gls{siui} }$. Note that \gls{siui} is a random binary vector},
  type=symbols
}

\newglossaryentry{cig}{%
  name={\ensuremath{\overline{c}_{\textnormal{IG}}}},
  description= {the expected number of molecules from transmitters of one group at receiver \gls{curRec} at sampling time $t = \gls{samplingTime}$},
  type=symbols
}

\newglossaryentry{cn}{%
  name={\ensuremath{c_{\textnormal{n}}}},
  description= {a random number which represents the number of noise molecules, so molecules from the background, at sampling time \gls{samplingTime} with $\gls{cn} \sim \pois{\gls{cmeann}} = \fpdf{\gls{cn}}$},
  type=symbols
}

\newglossaryentry{r}{%
  name={\ensuremath{r_{\mathrm{T}}}},
  description= {the number of received molecules with $\gls{r} = \gls{cs} + \gls{ciui} + \gls{cn} \sim \pois{\gls{ss}\gls{csmean} + \gls{siui}\gls{cmeaniui}^{\textnormal{T}} + \gls{cmeann}} = \fpdf{r \given \gls{ss}, \gls{siui}}$ and therefore $\fpdf{\gls{r} \given \gls{ss}} = \sum_{\gls{siui}}\fpdf{\gls{r} \given \gls{ss}, \gls{siui}} \underbrace{\fpdf{\gls{siui}}}_{\frac{1}{\gls{Riui}}}$},
  type=symbols
}

\newglossaryentry{threshold}{%
  name={\ensuremath{\xi}},
  description= {a scalar value. If the received signal $\gls{r} \geq \gls{threshold}$ the detector will decide in favor of a $1$, i.e. $  \hat{\gls{ss}} = 1$},
  type=symbols
}

\newglossaryentry{groupl}{%
  name={\ensuremath{l}},
  description= {group index \gls{groupl} of the interfering group},
  type=symbols
}

\newglossaryentry{nrInIUIGroupl}{%
  name={\ensuremath{N_{\textnormal{ILI},\gls{groupl}}}},
  description= {the number of interfering transmitters in group \gls{groupl}},
  type=symbols
}

\newglossaryentry{nrInIUIGroup}{%
  name={\ensuremath{N_{\textnormal{ILI}}}},
  description= {the number of interfering transmitters in group \gls{groupl}},
  type=symbols
}

\newglossaryentry{errorProb}{%
  name={\ensuremath{P_e}},
  description= {the number of interfering transmitters in group \gls{groupl}},
  type=symbols
}

\newglossaryentry{effectiveRate}{%
  name = {\ensuremath{R_{\textnormal{eff}}}},
  description = {the total achievable rate of a spatial multiplexing system as described in the paper},
  type= symbols
}

\newglossaryentry{spatialRate}{%
  name = {\ensuremath{R_{\textnormal{loc}}}},
  description = {the rate due to the density of transmitters in a spatial multiplexing system, which is therefore a function of the distance between transmitters or receivers, i.e. the spatial packing},
  type= symbols
}

\allowdisplaybreaks
\begin{document}
\bstctlcite{disable_url}
\title{Area Rate Efficiency in Multi-Link Molecular Communications}

\author{\IEEEauthorblockN{Lukas Brand, Sebastian Lotter, Vahid Jamali, Robert Schober, and Maximilian Schäfer}\thanks{This paper was presented in part at the ACM Nanoscale Computing and Communication Conference, 2021 \cite{Brand2021ARE}.}}

\maketitle
\nocite{Brand2021ARE}
\begin{abstract}
We consider a multi-link diffusion-based molecular communication (MC) system where multiple spatially distributed transmitter (TX)-receiver (RX) pairs establish point-to-point communication links employing the same type of signaling molecules. To exploit the full potential of such a system, an in- depth understanding of the interplay between the spatial link density and inter-link interference (ILI) and its impact on system performance is needed. In this paper, we consider a three-dimensional unbounded domain with multiple spatially distributed point-to-point non-cooperative transmission links, where both the TXs and RXs are positioned on a regular fixed grid. For this setup, we first derive an analytical expression for the channel impulse responses (CIRs) between the TXs and RXs in the system. Then, we derive the maximum likelihood (ML) detector for the RXs and show that it reduces to a threshold-based detector. Moreover, we derive an analytical expression for the corresponding detection threshold which depends on the statistics of the desired signal from the dedicated TX, the statistics of the MC channel, and the statistics of the ILI. We also provide a low-complexity suboptimal decision threshold. Furthermore, we derive an analytical expression for the bit error rate (BER) and the achievable rate of a single transmission link. Finally, we propose two new performance metrics, namely area rate efficiency (ARE) and area and time rate efficiency (ARTE), suitable for holistically evaluating spatially distributed multi-link MC systems. In particular, ARE and ARTE capture the tradeoff between transmission link density and achievable rate per link and the tradeoff between transmission link density, achievable rate per link, and inter-symbol interference (ISI), respectively. Hence, ARE and ARTE can be exploited to determine the optimal transmission link density for maximizing the throughput of the entire system.

\end{abstract}

\setlength{\belowdisplayskip}{2pt}
\setlength{\belowdisplayshortskip}{2pt}
\acresetall
\scaleSection
\section{Introduction}
\scaleSectionBelow
Molecular communication (MC)\acused{MC} is a bio-inspired paradigm, in which molecules are used to convey information. It is envisioned that \ac{MC} is an attractive alternative to electromagnetic wave based wireless communication in challenging environments such as sea water, pipe networks, etc. \cite{Farsad2016ACS}. However, the practical feasibility of \ac{MC} and the spectrum of future applications depend on whether it is possible to engineer \ac{MC} systems achieving high transmission reliability and large data rates. So far, the achievable data rates in \ac{MC} systems are comparatively low and typically only a single transmission link is considered, i.e., one \ac{TX}\acused{TXs} node releases molecules and one \ac{RX}\acused{RXs} node counts the received molecules. A common approach for increasing the throughput of \ac{MC} systems is to decrease the symbol duration, which however leads to \ac{ISI} \cite{Noel2014UENoiseISI, Tepekule2015ISIMItMC, Noel2014ISIMitByEnzymes}.

Besides the time dimension, the spatial dimension can be exploited to increase the system throughput. Hereby, choosing an appropriate spatial arrangement enables the efficient usage of this additional degree of freedom for system design. However, this option has received less attention in the \ac{MC} literature. Existing works exploit the spatial dimension for example in the context of \ac{MIMO} \ac{MC} systems \cite{Meng2012MIMOMC,Damrath2018ArGain,Huang2019SpatMod,Gursoy2019IndMod,Rouzegar2019MIMO} or for large-scale \ac{MC} systems with randomly distributed \ac{TXs} \cite{Deng2017MU3D, Zabini2019SDSecondOrder, Dissanayake2019InterferenceMitLarScal, Fang2021QuorumSensing}.
A \ac{MIMO} \ac{MC} system comprises one \ac{TX} and one \ac{RX}, but the \ac{TX} and the \ac{RX} are connected to multiple spatially distributed release and reception sites, respectively. Hence, well known techniques such as spatial modulation for encoding \cite{Huang2019SpatMod,Gursoy2019IndMod}, and selection combining and zero forcing for decoding \cite{Rouzegar2019MIMO, Meng2012MIMOMC} are applicable. Performance benefits of \ac{MIMO} \ac{MC} systems, such as diversity gain and spatial multiplexing gain, are discussed in \cite{Meng2012MIMOMC, Damrath2018ArGain}.
In all existing studies on \ac{MIMO} \ac{MC} systems, the considered number of release/reception sites is relatively low. To the best of the authors' knowledge, the largest system was investigated in \cite{Gursoy2019IndMod}, which studied an $8 \times 8$ system.
Unlike the studies on \ac{MIMO} \ac{MC}, large-scale \ac{MC} systems with an asymptotically large number of randomly distributed \ac{TXs} are investigated in \cite{Deng2017MU3D, Zabini2019SDSecondOrder, Dissanayake2019InterferenceMitLarScal, Fang2021QuorumSensing}. In \cite{Deng2017MU3D}, a swarm of randomly placed \ac{TXs} simultaneously transmit the same bit sequence to one \ac{RX}. Methods from stochastic geometry are utilized to analyze the received signal and the corresponding \ac{BER}. Stochastic geometry is also exploited in \cite{Zabini2019SDSecondOrder}, where the signal-to-noise ratio and the signal-to-interference ratio for randomly distributed \ac{MC} links are studied for both synchronous and asynchronous \ac{TXs}. The authors in \cite{Dissanayake2019InterferenceMitLarScal} extend the work of \cite{Deng2017MU3D, Zabini2019SDSecondOrder} by proposing an interference mitigation method based on error correction coding and a modulation scheme relying on two different types of molecules. In \cite{Fang2021QuorumSensing}, stochastic geometry is applied to study quorum sensing of randomly distributed bacteria. On the other hand, according to \cite{kim2008defined, boedicker2015microbial}, the spatial distribution of the bacteria impacts the efficiency of quorum sensing when multiple bacteria-based communication processes take place simultaneously. In this context, a system comprised of multiple different bacteria colonies can be modelled as a multi-link system. In such a system, quorum sensing enables communication within each colony and between different colonies due to the inter-link interference \cite[Sec. VII.]{boedicker2015microbial}, which facilitates cooperation, parasitism, and competition \cite{tan2015unraveling}. The latter is realized by causing intentional interference, i.e., some colonies use enzymes to destroy the quorum sensing signals of other colonies, see \cite[Fig. 7]{boedicker2015microbial} and \cite{rampioni2014art}. In \cite{silva2017signal, kim2008defined}, the authors show quantitatively that the interaction of spatially distributed colonies is distance dependent. In fact, in \cite{kim2008defined} synthetically designed bacteria colonies which are both competitive and cooperative are investigated. Here, cooperation is vital as the colonies possess complementary functions to re-mediate environmental contamination, but they compete for nutrients among the colonies. As shown in \cite[Fig. 3]{kim2008defined}, an optimal spatial separation between the colonies exist, which maximizes the viability of the colonies, i.e., a spatial distribution which balances the inter-colony competition and cooperation. These examples highlight the importance of studying the use of the spatial resource and developing performance metrics that allow the evaluation of multi-link \ac{MC} systems.

In order to gain a fundamental understanding of the benefits of exploiting the spatial dimension for \ac{MC}, in this paper, we study an \ac{MC} system with a large number of transmission links, i.e., an asymptotic regime in terms of the number of \ac{TX} and \ac{RX} nodes. In contrast to \cite{Deng2017MU3D, Zabini2019SDSecondOrder, Dissanayake2019InterferenceMitLarScal, Fang2021QuorumSensing}, where the \ac{TX} positions are \textit{random}, we assume \textit{deterministic} \ac{TX} and \ac{RX} positions, which allows us to reveal the impact of the choice of the \ac{TX} and \ac{RX} positions on the performance of the system.
We ask the following research question: \textbf{For given transmit and receive areas, accommodating multiple \ac{TXs} and \ac{RXs}, respectively, how densely should the \ac{TXs} and \ac{RXs} be deployed to maximize the information transmission rate?}
Similar to other resources such as transmission time and bandwidth in wireless communication, space is a limited resource, which is valuable in natural multi-link \ac{MC} systems, and is expected to also become valuable in synthetic systems. Hence, due to being limited, space allocation to different links needs to be optimized.
There exists a fundamental tradeoff between the achievable data rate of \textit{one} \ac{TX}-\ac{RX} link, which we refer to as link rate, and the link density. Increasing the density of \ac{TXs} and \ac{RXs} increases the number of transmission links per unit area. We refer to the number of links per unit area as spatial multiplexing rate with unit $[\si{ \per\square \meter}]$. However, the molecules released from different \ac{TXs} may cause \ac{ILI} at the \ac{RX} units. Therefore, increasing the density of \ac{TXs} and \ac{RXs} increases the \ac{BER} and consequently decreases the link rate. In order to analyze this tradeoff, we propose a new performance metric for multi-link \ac{MC} systems which we refer to as \ac{ARE}. The \ac{ARE} is the product of the spatial multiplexing rate and the link rate, and characterizes how efficiently the available \ac{TX} and \ac{RX} areas are used for information transmission and is given in terms of bits per unit area.

To systematically address the research question posed above, we study a \ac{3D} multipoint-to-multipoint \ac{MC} system comprising multiple non-cooperative, spatially distributed point-to-point transmission links and analyze the system in the asymptotic regime with a large number of \ac{TX} and  \ac{RX} nodes. As in conventional wireless communications, we define a location model for the positions of the \ac{TXs} and \ac{RXs} in the considered multi-link \ac{MC} setup \cite{Andrews2011CellNetwork,Nasri2016HexagonalNetwork}.
For wireless cellular networks, the two most common geometric models for the locations of base stations are fixed hexagonal and random grids, respectively \cite{Nasri2016HexagonalNetwork}. In both models, the positions of the users are assumed to be random. The authors in \cite{Andrews2011CellNetwork} show that, for real-world base station locations, one relevant performance metric, namely the signal-to-interference-plus-noise-ratio (SINR) experienced by the users, is upper bounded by the SINR of the users in an idealized grid network, and lower bounded by the SINR of the users in a random network.
While existing research on multi-link \ac{MC} systems is based on randomly positioned \ac{TXs} and \ac{RXs} \cite{Deng2017MU3D, Zabini2019SDSecondOrder, Dissanayake2019InterferenceMitLarScal, Fang2021QuorumSensing}, in this work, we assume an idealized, fixed, and predefined grid structure for the \ac{TXs} and \ac{RXs}. In particular, the \ac{TXs} and \ac{RXs} are arranged on a virtual \ac{TX} and \ac{RX} plane, respectively, cf. \Figure{fig:system_model}. Therefore, the model considered in this paper provides a new perspective on multi-link \ac{MC} systems. We evaluate the performance of the considered system by deriving an analytical expression for the \ac{ARE}. The main contributions of this work include:
\begin{enumerate}
  \item An analytical expression for the \ac{CIRs}\acused{CIR} of all \ac{TX}-\ac{RX} links for multipoint-to-multipoint molecule transmission via diffusion and uniform flow is derived assuming point \ac{TXs} and cylindrical, transparent \ac{RXs}.
  \item Based on the \ac{CIR} and a statistical model, mathematical expressions for the optimal detection threshold of a threshold detector, the \ac{BER}, the link rate, the spatial multiplexing rate, and finally the \ac{ARE} are derived in the asymptotic regime of a large number of \ac{TX}-\ac{RX} links.
  \item By analyzing the \ac{ARE}, we demonstrate a fundamental tradeoff between the spatial multiplexing rate and the \ac{ILI} dependent link rate on the overall performance of the considered multipoint-to-multipoint transmission system. Our analysis shows that there exists an optimal number of transmission links per unit area which maximizes the \ac{ARE}. We further show that an optimal link density also exists in the presence of \ac{ISI}, which is revealed by analyzing an extension of the \ac{ARE} denoted as \ac{ARTE}.
  \item We show that the optimal transmission link density, which maximizes the \ac{ARE}, depends on the number of released molecules and the diffusion coefficient of the molecules. Furthermore, our results reveal that the background noise concentration and the choice of the grid structure have negligible impact on the optimal link density.
\end{enumerate}

The \ac{ARE} was introduced in the conference version \cite{Brand2021ARE} of this paper. Compared to \cite{Brand2021ARE}, the system model in this paper also accounts for background noise molecules. Furthermore, in addition to the hexagonal grid for the \ac{TX} and \ac{RX} positions considered in \cite{Brand2021ARE}, we also study a square grid. Moreover, a suboptimal threshold based detector is developed, which enables low complexity threshold detection for cases in which the optimal threshold based detector might not be feasible. In addition, the impact of \ac{ISI} on the optimal link density is investigated.

The remainder of this paper is organized as follows. \Section{sec:model} introduces the system model and performance metrics for evaluation of the considered multi-link system. In \Section{sec:math}, an analytical expression for the link \ac{CIR} is derived and a statistical model for the considered \ac{MC} system is provided. The threshold character of the maximum likelihood detector is unveiled in \Section{sec:performance}, where also the resulting \ac{BER} is derived. In \Section{sec:evaluation}, the \ac{BER} and \ac{ARE} are studied and compared to particle-based and Monte Carlo simulations. Finally, in \Section{sec:conclusion}, our main conclusions are summarized.

\scaleSection
\section{System Model}\label{sec:model}
\scaleSectionBelow
In this section, we discuss the general arrangement of the \ac{MC} system, the fixed and predefined grid structure, and the model for modulation, propagation, and reception used in this paper. Moreover, we present different metrics for characterizing the performance of information transmission in the considered system. Hereby, we introduce the \ac{ARE} as a new performance metric for evaluation of the performance of multi-link \ac{MC} systems. Finally, extending the definition of the \ac{ARE}, we define the \ac{ARTE} as a performance metric for cases, where \ac{ISI} occurs.

\scaleSubsection
\subsection{General Arrangement}\label{general_setup}
\scaleSubsectionBelow

\begin{figure}[!tbp]
  \begin{minipage}[t]{1\columnwidth}
    \includegraphics[width=1\textwidth]{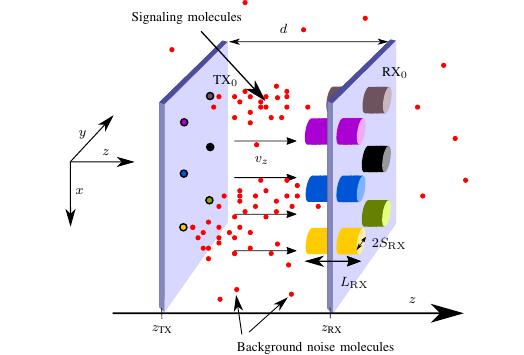}
    \caption{\acs{MC} system model with multiple, independent point-to-point transmission links (highlighted by different colors for \acs{TXs} and \acs{RXs}), i.e., each TX intends to communicate with only one dedicated RX. The point \acs{TXs} are aligned within a predefined grid in a \acs{TX} plane, shown in blue on the left hand side, and release molecules (red dots) using ON-OFF keying modulation. The release process at the TXs is assumed to be time synchronized. The molecules propagate by Brownian motion and uniform flow and are detected by transparent cylindrical \acs{RXs} which are aligned in a \acs{RX} plane, depicted in blue on the right hand side, in the same grid as the \acs{TXs}. Both planes are virtual, i.e., transparent and of infinite extent. Additionally, background noise molecules (red dots, labeled accordingly) exist, which are not distinguishable from signaling molecules by the RX.}\label{fig:system_model}
  \end{minipage}
\end{figure}
We consider a \ac{3D} unbounded \ac{MC} environment with an infinite number of point \ac{TXs} and an infinite number of transparent \ac{RXs}\footnote{The transparent, molecule counting RX model is widely used in the MC literature, see \cite{jamali2016channel,meng2014receiver, kilinc2013receiver}, mainly due to its passivity, i.e., such \ac{RXs} do not impair molecule movement. For complex \ac{MC} systems, such as the one considered in this work, transparent \ac{RXs} facilitate the analysis.}, cf. \Figure{fig:system_model}. The \ac{TXs} and \ac{RXs} are located in the \mbox{$xy$}-plane at $z=z_{\mathrm{TX}}$ and at $z=z_{\textnormal{RX}}$, respectively. Both planes are virtual (i.e., they are transparent), have infinite width and height (i.e., they extend infinitely in the $x$ and $y$ directions), and are placed $\gls{d} = z_{\textnormal{RX}} - z_{\mathrm{TX}} $ apart. In the following, these planes are referred to as TX plane and RX plane, respectively. Each \ac{TX} in the TX plane is paired with the closest \ac{RX} in the RX plane, i.e., each \ac{TX} wants to communicate with only one dedicated \ac{RX}. We denote the individual \ac{TXs} and \ac{RXs} as $\mathrm{TX}_i$ and $\mathrm{RX}_j$, $i, j \in \mathbb{N}$, respectively. We note that the number of \ac{TXs} is equal to the number of \ac{RXs}. Furthermore, we assume uniform flow in $z$ direction with velocity $\gls{flow}$. The chosen setup ensures error-free transmission if only flow is present and impairing factors such as diffusion and noise sources are absent. Of course, due to the use of small-sized molecules, diffusion always occurs in practice. Hence, the random nature of diffusion degrades the performance of the considered transmission links.

\scaleSubsection
\subsection{Grid Structures}\label{grid_section}
\scaleSubsectionBelow

\begin{figure}
     \begin{minipage}[t]{1\columnwidth}
       \centering
       \includegraphics[width=1\textwidth]{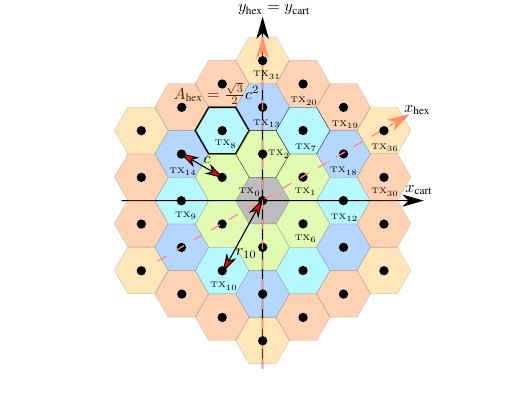}
      \caption{The \ac{TX} positions are shown as black dots within the virtual hexagonal grid with axes $x_{\textnormal{hex}}$ and $y_{\textnormal{hex}}$; the axes of the Cartesian coordinate system $x_{\textnormal{cart}}$ and $y_{\textnormal{cart}}$ are included as a reference. The distance $\gls{r0minHex}$ between two adjacent \ac{TXs} and, as an example, the distance $r_{10}$ between the reference \ac{TX} in the center and $\ac{TX}_{10}$ as an example are depicted by arrows. Multiple \ac{TXs} having the same distance to \gls{RX0} are conceptionally grouped into rings of interferers which are highlighted with the same grid fill color, e.g., the cells of $\ac{TX}_1-\ac{TX}_6$, $\ac{TX}_7-\ac{TX}_{18}$, and $\ac{TX}_{19}-\ac{TX}_{36}$ are in green, blue, and orange, respectively. This eases the identification of the structure composed by hexagons.}
       \label{graphic:System_model:Grid_hex}
     \end{minipage}
     \begin{minipage}[t]{1\columnwidth}
       \centering
       \includegraphics[width=1\textwidth]{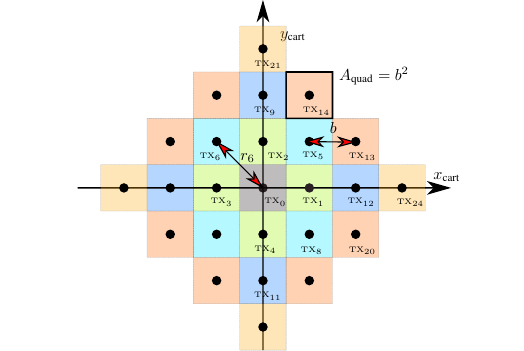}
       \caption{The \ac{TX} positions are shown as black dots within the virtual square grid with the Cartesian coordinate system $x_{\textnormal{cart}}$ and $y_{\textnormal{cart}}$ as axes. The distance $b$ between two adjacent \ac{TXs} and, as an example, the distance $r_{6}$ between the reference \ac{TX} in the center and $\ac{TX}_{6}$ are depicted by arrows. Multiple \ac{TXs} having the same distance to \gls{RX0} are conceptionally grouped into rings of interferers which are highlighted with the same grid fill color, e.g., the cells of $\ac{TX}_1-\ac{TX}_4$, $\ac{TX}_5-\ac{TX}_{12}$, and $\ac{TX}_{13}-\ac{TX}_{24}$ are in green, blue, and orange, respectively. This eases the identification of the structure composed by squares.}
       \label{graphic:System_model:Grid_quad}
     \end{minipage}
\end{figure}

We assume that the TX and RX planes are subdivided into an infinite number of equally shaped cells, respectively, resulting in a cellular grid. We consider two different cell shapes, namely hexagons and squares, i.e., both the \ac{TX} and \ac{RX} planes are either subdivided into a hexagonal grid or into a square grid, respectively, cf. \Figures{graphic:System_model:Grid_hex}{graphic:System_model:Grid_quad}. The hexagonal grid is widely used for the design of cellular architectures, e.g., in wireless communication networks \cite{Andrews2011CellNetwork,Nasri2016HexagonalNetwork}. Furthermore, hexagons are the shape closest to a circle that can form a continuous grid, and therefore, guarantee the densest packing of non-overlapping cells. The square grid is suboptimal in terms of packing, however, it is simple to describe in Cartesian coordinates. Each point \ac{TX} and the center of each \ac{RX} are located in a cell center. The center points of the hexagons and squares can be given in terms of an offset coordinate system $x', y'$\footnote{Using an offset coordinate system is an approach to uniquely label cells in hexagonal grids by integer tuples of $x', y'$, which simplifies the follow-up geometric analyses, e.g., the derivation of distances. Finally, the results obtained for the offset coordinate system can be related to a different coordinate system, e.g., the Cartesian coordinate system, by coordinate transformation methods.}, as defined below, and the Cartesian coordinate system $x, y$, respectively. The center points are identical for the \ac{TX} plane and the \ac{RX} plane\footnote{Perfect alignment of the transmission and reception sites is an assumption used also in the existing literature on distributed MC systems \cite{Koo2016MoMIMO, Gursoy2019IndMod, angjo2023molecular}. In addition, for conventional wireless communication networks, systems where the base stations are organized according to an idealized grid have been shown to provide performance upper bounds for practical wireless cellular systems \cite{Andrews2011CellNetwork}. Hence, the \ac{MC} system considered in this work may also provide a performance bound for \ac{MC} systems with non-aligned \ac{TXs} and \ac{RXs}. Proving this conjecture is an interesting topic for future work.}.
The Euclidean distances between the center points of two adjacent hexagonal and square cells are given by $\gls{r0minHex}$ and $b$, respectively, see \Figures{graphic:System_model:Grid_hex}{graphic:System_model:Grid_quad}. The distance of the center points between \gls{TX0} with $(x'_{\textnormal{TX}_0},y'_{\textnormal{TX}_0}) = (0,0)$ and $(x_{\textnormal{TX}_0},y_{\textnormal{TX}_0}) = (0,0)$ and $\mathrm{TX}_{i}$ with $(x'_{\textnormal{TX}_i},y'_{\textnormal{TX}_i})$ and $(x_{\textnormal{TX}_i},y_{\textnormal{TX}_i})$ is given as $\gls{r0minHex} \, \sqrt{ x'^2_{\textnormal{TX}_i}  + y'^2_{\textnormal{TX}_i}  + x'_{\textnormal{TX}_i} \; y'_{\textnormal{TX}_i} } $ and $b \sqrt{x_{\textnormal{TX}_i}^2 + y_{\textnormal{TX}_i}^2}$, respectively. The offset coordinate system $x', y'$ is related to the Cartesian coordinate system $x,y$ by $x = \gls{r0minHex} \, \frac{\sqrt{3}}{2} x', y = \gls{r0minHex} \, (y' + \frac{1}{2} x')$. The area size of one hexagon and one square is given by $A_\mathrm{hex} = \frac{\sqrt{3}}{2} \gls{r0minHex}^2$ and $A_\mathrm{quad} = b^2$, respectively. To enable a fair comparison between both grid structures, the cell center distances $b$ and $\gls{r0minHex}$ are chosen such that the area size of the hexagon cell equals the area size of the square cell, i.e., $A_\mathrm{hex} = A_\mathrm{quad}$, which is ensured by choosing $ b =  \gls{r0minHex}  \, \sqrt{\frac{\sqrt{3}}{2}}$.
The considered arrangement ensures that for $\textnormal{TX}_i$, the distance to $\textnormal{RX}_i$ is shorter than the distance to any other $\textnormal{TX}_j$, $i \neq j$. Moreover, without loss of generality, in the following, we exemplary analyse the point-to-point transmission link between \gls{TX0} and \gls{RX0}, which are both at the origin of the \mbox{$x$-$y$ plane}, see \Figures{graphic:System_model:Grid_hex}{graphic:System_model:Grid_quad}. All other \ac{TXs} act as interferers for $\gls{TX0}$. All results obtained for this \ac{TX}-\ac{RX} pair are also valid for any other TX-RX pair due to the symmetry of the considered grids and the infinite extent of the \ac{TX} and \ac{RX} planes. While practical \ac{MC} systems have finite dimensions, considering a system with infinite dimensions allows us to gain fundamental insights for system design. Furthermore, considering infinite \ac{TX} and \ac{RX} planes provides an upper bound for the \ac{ILI} for systems with a finite number of \ac{TXs} and \ac{RXs} as it constitutes the worst-case scenario in terms of \ac{ILI}.

\scaleSubsection
\subsection{Modulation, Propagation, and Reception}\label{sys_model:mod_prop_rec}
\scaleSubsectionBelow

At the beginning of each symbol interval $l \in \mathbb{N}$, each point $\mathrm{TX}_i$ releases a fixed non-zero number of molecules, \gls{nrMol}, or zero molecules, representing binary symbols $s_{i,l} = 1$ and $s_{i,l} = 0$, respectively, i.e., the \ac{TXs} use \acf{OOK} modulation \cite{Jamali2019ChannelMF}. Since the focus of this paper is the spatial dimension, we assume, for now, that \ac{ISI} is negligible which is a valid assumption if the symbol durations are sufficiently large\footnote{We note that for some \ac{MC} systems large symbol intervals may not be optimal for maximizing data throughput, of course. Instead, small symbol intervals may be preferable which leads to \ac{ISI}. However, including ISI into the system model proposed in this work renders the mathematical analysis infeasible. Thus, we neglect \ac{ISI} in our analytical work, but provide insight into the impact of \ac{ISI} based on computer simulations in \Section{section:evaulation:ISI:results}.}. Thus, the transmitted symbols are orthogonal in time, and therefore, we omit symbol interval index $l$ in the following. We assume that binary values $0$ and $1$ are equiprobable, i.e., $\prob{s_i = 1} = 0.5 = \prob{s_i = 0}$. The molecules emitted by different \ac{TXs} are of the same type, and the \ac{TXs} are synchronized, i.e., they transmit at the same time instants\footnote{Time synchronization is a common assumption in the MC literature \cite{Fang2021QuorumSensing, Gursoy2019IndMod, Koo2016MoMIMO} and can be achieved by adopting existing \ac{MC} synchronization techniques \cite{shahmohammadian2013blind, lin2016time, luo2016offset, jamali2017symbol}.}. We assume that molecules do not interact with each other, i.e., the release, propagation, and reception of different molecules are independent, respectively. The propagation of the molecules released by the \ac{TXs} is affected by diffusion, characterized by diffusion coefficient \gls{diffusion}, and uniform flow in $z$-direction, characterized by flow velocity $\gls{flow}$. We neglect additional effects influencing the propagation of molecules such as external forces, turbulences in the flow profile, and degradation.
Each transparent \ac{RX} counts the number of molecules within its volume \gls{VRX} at a fixed sampling time \gls{samplingTime}. The sampling time \gls{samplingTime} at \gls{RX0} is chosen such that it coincides with the time instant where the \ac{CIR} between \gls{TX0} and \gls{RX0} has its peak. The \ac{RXs} are identical cylinders with radius $\gls{SRX}$ and length $\gls{LRX}$, i.e., $\gls{VRX} = S_{\mathrm{RX}}^2 \pi \gls{LRX}$, see \Figure{fig:system_model}. We chose cylindrical receivers as this allows an independent scaling of the receiver size in the \mbox{$xy$}-plane via \gls{SRX} and along the $z$-axis via \gls{LRX}\footnote{We note that spherical \ac{RXs} are considered for reference in \Section{sec:CIR_verification}.}. Furthermore, the circular cross-section of the \ac{RX} fills both the hexagonal grid and the square grid to a great extent if \gls{SRX} is chosen large.

\scaleSubsection
\subsection{Performance Metrics for Multi-Links \ac{MC} Systems}\label{Performance_metrics}
\scaleSubsectionBelow
In this section, we define the single link rate, the spatial multiplexing rate, and the \ac{ARE}, for performance evaluation of multi-link \ac{MC} systems. Finally, we define the \ac{ARTE} for \ac{MC} systems employing short symbol intervals such that \ac{ISI} occurs. The \ac{ARTE} extends the definition of the \ac{ARE}, and therefore, it is applicable for all multi-link \ac{MC} systems.
\scaleSubsubsection
\subsubsection{Link Rate}
\scaleSubsubsectionBelow
We model the point-to-point transmission link as a binary channel with achievable data rate \cite[Eq. (9.7)]{Mackay2003information}
\begin{align}
  R_{\textnormal{SISO}} = \mathrm{I}(\gls{ss};\gls{ssHat}) = \mathrm{H}(\gls{ssHat}) - \mathrm{H}(\gls{ssHat} \given \gls{ss}) \;,
  \label{eq:performacne_user_rate}
\end{align}
as \ac{OOK} is a binary modulation scheme, i.e., $\gls{ss}, \gls{ssHat} \in \{0,1\}$.
We refer to $R_{\textnormal{SISO}}$ as link rate.
$\gls{ss}$ and $\gls{ssHat}$ correspond to the channel input symbol and the channel output symbol, respectively, and $\mathrm{I}(\cdot;\cdot)$ and $\mathrm{H}(\cdot)$ denote the mutual information and the entropy function, respectively, where \cite[Eq. (2.35)]{Mackay2003information} $\mathrm{H}(\gls{ssHat}) = -\frac{1}{2}\left[\left(\overline{p}+q\right) \log_2\left(\frac{1}{2} \left(\overline{p}+q\right)\right) + \left(\overline{q}+p\right) \log_2\left(\frac{1}{2} \left(\overline{q}+p\right)\right)\right]$ and \cite[Eq. (8.4)]{Mackay2003information} $\mathrm{H}(\gls{ssHat} \given \gls{ss}) = -\frac{1}{2} \left(\left(\overline{p}\right) \log_2\left(\overline{p}\right) + p \log_2\left(p\right) + q \log_2\left(q\right) + \left(\overline{q}\right) \log_2\left(\overline{q}\right)\right)$  with $\overline{p} = 1-p$ and $\overline{q} = 1-q$.
Here, $q = \prob{\gls{ssHat} = 0 \given \gls{ss} = 1}$ and $p = \prob{\gls{ssHat} = 1 \given \gls{ss} = 0}$ denote the error probability for $\gls{ss} = 1$ and $\gls{ss} = 0$, respectively.

\scaleSubsubsection
\subsubsection{Spatial Multiplexing Rate}
\scaleSubsubsectionBelow
We define the spatial multiplexing rate \gls{spatialRate} as the number of transmission links per unit area, i.e., $ \gls{spatialRate} = \frac{1}{A_{\textnormal{cell}}}$ with unit $[\si{ \per\square \metre}]$, where $A_{\textnormal{cell}}$ is the area size reserved for one TX-RX pair, i.e., one hexagon or one square. Hence, the smaller $A_{\textnormal{cell}}$ is, the larger is the spatial multiplexing rate.
\scaleSubsubsection
\subsubsection{Area Rate Efficiency}
\scaleSubsubsectionBelow
We define the \ac{ARE} based on $R_{\textnormal{SISO}}$ and $\gls{spatialRate}$ as follows
\begin{align}
  \mathrm{ARE} &= R_{\textnormal{SISO}} \gls{spatialRate} =   \frac{1}{A_{\textnormal{cell}}} [\mathrm{H}(\gls{ssHat}) - \mathrm{H}(\gls{ssHat} \given \gls{ss}) ] \;,
\label{eq:performance:effRate}
\end{align}
with unit $[\mathrm{bit }\,\si{ \per\square\metre}]$. Note that the two rates, $R_{\textnormal{SISO}}$ and $\gls{spatialRate}$, exhibit a different dependence on the density of the \ac{TXs} and \ac{RXs}. In particular, increasing the density of the TX-RX pairs increases $\gls{spatialRate}$ but decreases $R_{\textnormal{SISO}}$ as the \ac{BER} increases, and vice versa. Hence, the \ac{ARE} reflects both the single link performance in terms of $R_{\textnormal{SISO}}$ \textit{and} the area usage efficiency in terms of $\gls{spatialRate}$. Therefore, the \ac{ARE} provides a useful performance metric to gain insight into how efficiently given \ac{TX} and \ac{RX} areas are exploited for the maximization of the overall information transmission.

  \scaleSubsubsection
  \subsubsection{Area and Time Rate Efficiency}\label{sss:achievable_rate}
  \scaleSubsubsectionBelow
  We define the \ac{ARTE} based on $R_{\textnormal{SISO}}'$, $\gls{spatialRate}$, and symbol duration $T_{\mathrm{sym}}$ as follows
  \scaleAlign
  \begin{align}
    \mathrm{ARTE} &= R_{\textnormal{SISO}}' \gls{spatialRate} \frac{1}{T_{\mathrm{sym}}}  =   \frac{1}{A_{\textnormal{cell}}} [\mathrm{H}(\gls{ssHat}) - \mathrm{H}(\gls{ssHat} \given \gls{ss}) ] \frac{1}{T_{\mathrm{sym}}} \;,
  \label{eq:performance:achievableRate}
  \end{align}
with unit $[\mathrm{bit }\,\si{ \per\square\metre \per\second}]$. Here, $R_{\textnormal{SISO}}'$ denotes the link rate in the presence of \ac{ISI}. From \Equation{eq:performance:achievableRate} we observe that, unlike the \ac{ARE}, the \ac{ARTE} incorporates the symbol duration. Hence, the \ac{ARTE} provides insight into how efficiently the given \ac{TX} and \ac{RX} areas are exploited in a given time interval. The \ac{ARTE} in \Equation{eq:performance:achievableRate} enables a comprehensive analysis of multi-link \ac{MC} systems, as both time and area usage in conjunction with the impact of \ac{ILI} and \ac{ISI} are considered. However, deriving analytical expressions for the \ac{BER} and therefore the \ac{ARTE} for this case appears to be infeasible due to the complex joint \ac{ISI} and \ac{ILI} statistics. In fact, the overall interference comprises \ac{ISI} from \gls{TX0}, interference from the other \ac{TXs}, and \ac{ISI} from the other \ac{TXs}. Hence, we do not pursue analytical expressions for the \ac{ARTE} and instead analyze the \ac{ARTE} based on Monte Carlo simulations in \Section{section:evaulation:ISI:results}.

To be able to analyze the considered multi-link system in terms of its \ac{ARE} and \ac{ARTE}, in the following, we first analyze the relevant \ac{MC} channels, detection design, and transmission errors, respectively.

\scaleSection
\section{Analytical Channel Model}\label{sec:math}
\scaleSectionBelow
In this section, we first derive an expression for the \ac{CIR} of all transmission links that involve receiver \gls{RX0}. We note that the derived \ac{CIR} expression is also valid for all other \ac{TX}-\ac{RX} pairs due to the symmetries in the considered system model. Then, we analyze the distributions of the information, the interfering, and the background noise molecules, respectively.
\scaleSubsection
\subsection{Channel Impulse Response}\label{math_sec:cir}
\scaleSubsectionBelow
$\textnormal{CIR}_{\gls{curTrans}}(t)$ denotes the probability to observe one molecule at $\gls{RX0}$ at time $t$ which was released by $\ac{TX}_{\gls{curTrans}}$ at position\footnote{Note that $z_{\textnormal{TX},i} = z_{\textnormal{TX}}$, $\forall i$, holds as all \ac{TXs} are located in the \ac{TX} plane.} $\boldsymbol{p}_{\gls{curTrans}} = (x_{\textnormal{TX}_i}, y_{\textnormal{TX}_i}, z_{\textnormal{TX}})$ and at time $t_{0} = 0$. In the following proposition, we provide an analytical expression for $\textnormal{CIR}_{\gls{curTrans}}(t)$ for the considered unbounded environment with constant advection along the $z$-axis with velocity $\gls{flow}$ and diffusion coefficient $\gls{diffusion}$.
\begin{prop}\label{prop:cir}
The expected number of received molecules at \gls{RX0}, centered at $(0,0, z_{\mathrm{RX}})$ with radius \gls{SRX} and extending in $z$-direction from $z_{\mathrm{S}} = z_{\mathrm{RX}} - \frac{\gls{LRX}}{2} $ to $z_{\mathrm{E}} = z_{\mathrm{S}} + \gls{LRX} = z_{\mathrm{RX}} + \frac{\gls{LRX}}{2}$, due to the release of one molecule by $\ac{TX}_{\gls{curTrans}}$ at time $t_0 = 0$, is given by
\scaleAlign
  \begin{align}
  &\textnormal{CIR}_{\gls{curTrans}}(t) \nonumber \\
  &= \frac{1}{2} \left(\erf\left(\frac{z_{\textnormal{TX}} + \gls{flow} t - z_{\mathrm{S}}}{\sqrt{4 D t}}\right)- \erf\left(\frac{z_{\textnormal{TX}} +\gls{flow} t - z_{\mathrm{E}}}{\sqrt{4 \gls{diffusion} t}}\right)\right) \nonumber \\
  & \qquad \times \exp\left(-\frac{r_{\gls{curTrans}}^2}{4 \gls{diffusion} t}\right) \sum_{k=0}^{k_{\mathrm{max}} = \infty} \frac{(\frac{r_{\gls{curTrans}}^2}{4 \gls{diffusion} t})^k}{(k!)^2} \gamma\left(k+1,\frac{\gls{SRX2}}{4 \gls{diffusion} t}\right) \;,
  \label{eq:math_ir_general}
  \end{align}
  where $r_{\gls{curTrans}}^2 = x^2_{\textnormal{TX}_i} + y^2_{\textnormal{TX}_i}$. Furthermore, $\erf(x)$ and $\gamma(a,x)$ denote the Gaussian error function and the lower incomplete Gamma function, respectively.
\end{prop}
\begin{IEEEproof}
  Please refer to \Appendix{section:appendix:CIRProof}.
\end{IEEEproof}
Eq. \Equation{eq:math_ir_general} shows that the expected number of molecules received from $\textnormal{TX}_i$ depends on the \ac{TX} position characterized by $r_{\gls{curTrans}}$. We observe that $\textnormal{CIR}_{\gls{curTrans}}(t) \rightarrow 0$ for \ac{TXs} that are far away from $\textnormal{RX}_0$, because, in this case, $r_{\gls{curTrans}} \rightarrow \infty$. All other parameters in \Equation{eq:math_ir_general}, e.g., the $z$-position of the \ac{TXs}, the diffusion coefficient $\gls{diffusion}$, and the flow velocity \gls{flow} are identical for all \ac{TXs}.

  \begin{prop}\label{prop_k_max_test}
    For evaluation of \Equation{eq:math_ir_general}, we can ensure that the truncation error for $\textnormal{CIR}_{\gls{curTrans}}(\gls{samplingTime})$, $i>0$, is below a predefined fraction $\eta$ of $\textnormal{CIR}_{0}(t = \gls{samplingTime})$\footnote{Note that for \gls{TX0}, $r_0 = 0$ follows and $\textnormal{CIR}_{0}(t = \gls{samplingTime}) = \frac{1}{2} \left(\erf\left(\frac{z_{\textnormal{TX}} + \gls{flow} \gls{samplingTime} - z_{\mathrm{S}}}{\sqrt{4 D \gls{samplingTime}}}\right)- \erf\left(\frac{z_{\textnormal{TX}} +\gls{flow} \gls{samplingTime} - z_{\mathrm{E}}}{\sqrt{4 \gls{diffusion} \gls{samplingTime}}}\right)\right) \left(1-\exp\left(-\frac{\gls{SRX2}}{4 \gls{diffusion} \gls{samplingTime}}\right)\right)$ can be evaluated exactly.} when setting $k_{\mathrm{max}}$ to a finite number $k_{\mathrm{max}} = k'$, i.e., for given $\eta$ there exist $k'$ such that
    \scaleAlign
    \begin{align}
      &\quad \textnormal{CIR}_{\gls{curTrans}}(t = \gls{samplingTime}) \given{_{k_{\mathrm{max}} = \infty}} \, - \, \textnormal{CIR}_{\gls{curTrans}}(t = \gls{samplingTime}) \given{_{k_{\mathrm{max}} = k'}} \nonumber \\
      & \quad \quad < \,\eta \; \textnormal{CIR}_{0}(t = \gls{samplingTime}) \;.
    \end{align}
  \end{prop}
  \begin{IEEEproof}
    Please refer to \Appendix{section:appendix:kmax_truncation}.
  \end{IEEEproof}
%

\scaleSubsection
\subsection{Statistical Model}\label{Sec:statistic_model}
\scaleSubsectionBelow

The total number of observed molecules, $\gls{r}$, within \gls{RX0} comprises the information conveying molecules originating from \gls{TX0}, $\gls{cs}$, molecules received from other \ac{TXs}, $\gls{ciui}$, which constitute \ac{ILI}, and background noise molecules, \gls{cn}. Hence, the number of observed molecules at sampling time $\gls{samplingTime}$ at \gls{RX0} is given by
\scaleAlign
\begin{align}
  \gls{r} = \gls{cs} + \gls{ciui} + \gls{cn}\;.
  \label{eq:system_model_received_molecules}
\end{align}

We now discuss the underlying models and associated distributions of $\gls{cs}$, $\gls{ciui}$, and \gls{cn}.
\scaleSubsubsection
\subsubsection{Information Molecules}\label{ssSec:IM}
\scaleSubsubsectionBelow
\gls{nrMol} information carrying molecules intended for \gls{RX0} are released by \gls{TX0} for signaling $\gls{ss}=1$ at time $t=0$. At time instant $t_\mathrm{m}$, each of these molecules is received with probability $\textnormal{CIR}_{0}(t=t_\mathrm{m})$ at \gls{RX0}. The expected (non-normalized) number of observed information molecules at sampling time $\gls{samplingTime}$ is given by $ \gls{csmean} =  \gls{nrMol} \; \textnormal{CIR}_{0}(\gls{samplingTime}).$
We model the received particle statistics by a Poisson distribution, i.e.,
\scaleAlign
\begin{align}
  \gls{cs} \sim \pois{\gls{ss}\gls{csmean} } = \fpdf[s]{\gls{cs} \given \gls{ss} } \;,
  \label{equation:math_sec:ss_pois}
\end{align}
which is a valid approximation for molecule counting \ac{RXs} if the number of released molecules is large \cite{Jamali2019ChannelMF}. Here, $\fpdf[s]{\gls{cs} \given \gls{ss} }$ denotes the distribution of $\gls{cs}$ conditioned on \gls{ss}.


\scaleSubsubsection
\subsubsection{Interference Molecules from other Transmitters}\label{ssSec:IUIMol}
\scaleSubsubsectionBelow
Besides the information molecules, molecules emitted by $\mathrm{TX}_1, \mathrm{TX}_2, \mathrm{TX}_3, \dots, \mathrm{TX}_{\gls{nrTrans}-1}$, i.e., the \ac{TX}s belonging to other TX-RX pairs, are received as \ac{ILI} at $\gls{RX0}$, where $\gls{nrTrans} \rightarrow \infty$ denotes the number of transmitters. Due to the infinite extent of the \ac{TX} plane, infinitely many interferers exist in principle.
We assume that the interfering molecules are not distinguishable from the information molecules. Hence, \ac{ILI} degrades the detection of the signal of \gls{TX0} at \gls{RX0} \cite{Noel2014UENoiseISI}. We note that $\ac{TX}_{\gls{curTrans}}$, $\gls{curTrans} \neq 0$, causes interference at \gls{RX0} only for $s_{\gls{curTrans}} = 1$ as no molecules are released for $s_{\gls{curTrans}} = 0$. Hence, in addition to the Brownian motion of the molecules, the random transmit sequence at $\ac{TX}_{\gls{curTrans}}$ introduces randomness into the received signal. So, for a comprehensive statistical analysis, the joint statistics of the transmit sequence at all \ac{TX}s as well as the statistics of the molecule movement have to be taken into account.

From the perspective of \gls{RX0}, the expected number of received molecules at sampling time $t = \gls{samplingTime}$ can be collected in vector $\gls{cmeaniui} = [\overline{c}_{1}, \overline{c}_{2}, \dots, \overline{c}_{\gls{nrTrans}-1}] $ with $ \overline{c}_{i} =  \gls{nrMol} \; \textnormal{CIR}_{i}(t=\gls{samplingTime})$, assuming molecule releases at all interferers.
Similar to \Equation{equation:math_sec:ss_pois}, the received number of molecules from each interferer can be characterized by a Poisson distribution \cite{Jamali2019ChannelMF}, as the molecule releases of the different \ac{TX}s are independent of each other. As \gls{RX0} counts all molecules, independent of their origin, the sum over all Poisson distributed interfering molecules follows again a Poisson distribution, i.e.,
\scaleAlign
\begin{align}
   \gls{ciui} \sim \pois{\gls{siui}\gls{cmeaniuiTrans} } = \fpdf[ILI]{\gls{ciui} \given \gls{siui} } \;,
   \label{equation:math_sec:siui_pois}
\end{align}
where $\fpdf[ILI]{\gls{ciui} \given \gls{siui} }$ denotes the distribution of \gls{ciui} conditioned on vector $\gls{siui}$, which contains the symbols emitted by the interfering \ac{TXs}, i.e., $\gls{siui} = [s_1, s_2, \dots, s_{\gls{nrTrans}-1}]$. There are $\gls{Riui} = 2^{\gls{nrTrans}-1}$ possible realizations of \gls{siui}, i.e., different \ac{ILI} states, which are equiprobable due to the equiprobable binary transmission symbols.
Hence, $\gls{Riui}$ grows exponentially in $\gls{nrTrans}$ and we assumed $\gls{nrTrans} \rightarrow \infty$ in our system model.
However, the \ac{ILI} can be accurately approximated by truncating \gls{nrTrans} to a finite number, i.e., by taking into account only a finite number of interfering \ac{TXs}. The \ac{ILI} is mainly characterized by the strongest interferers, which correspond to the \ac{TXs} closest to \gls{TX0}, as will be confirmed via simulations in \Section{section:evaluation:BER_over_threshold}. As can be observed from \Figure{graphic:System_model:Grid_hex}, the indices of the \ac{TXs} are chosen such that, for increasing index $i$, the distance between \gls{TX0} and $\mathrm{TX}_{\gls{curTrans}}$ is monotonically non-decreasing in \gls{curTrans}. Therefore, the \textit{strongest} interferers are always included if \gls{nrTrans} is truncated to a finite number. We note that the actual number of \ac{TXs} necessary to accurately approximate the system behavior depends on the values of the system parameters, see \Section{eval:BER_ARE_Results}.
\scaleSubsubsection
\subsubsection{Background Noise Molecules}\label{section:math_section:Background_noise}
\scaleSubsubsectionBelow
Besides information molecules and molecules from interfering \ac{TX}s, we also account for background noise molecules, which are of the same type as the signaling molecules. These molecules may originate from far away noise sources. We assume that these molecules are uniformly distributed in space with constant concentration $C_{\textnormal{noise}}$. We statistically model the \textit{reception} of the background noise molecules by a Poisson distribution, i.e.,
\scaleAlign
\begin{align}
  \gls{cn} \sim \pois{\gls{cmeann}} = \fpdf[n]{\gls{cn}},
  \label{equation:math_sec:n_pois}
\end{align}
which is a valid approximation based on the law of rare events \cite{Jamali2019ChannelMF}. Here, $\fpdf[n]{\gls{cn}}$ denotes the distribution of $\gls{cn}$, and \gls{cmeann} is given by $\gls{cmeann} = C_{\textnormal{noise}} \gls{VRX}$.

\scaleSection
\section{Symbol Detection and BER Performance Analysis}\label{sec:performance}
\scaleSectionBelow
In this section, we derive the optimal \ac{ML} decision rule and the \ac{BER} of one TX-RX pair.
\scaleSubsection
\subsection{Optimal ML Detector}\label{math_sec:ssMLDet}
\scaleSubsectionBelow
For detection, we assume all \ac{CIR}s in the system are known but the activity of the interfering \ac{TXs} is only statistically known. Therefore, the \ac{ML} estimate, \gls{ssHat}, is given by
\scaleAlign
\begin{align}
\gls{ssHat} &= \underset{\gls{ss} \in \{0,1\}}{\text{argmax}} \; \fpdf[\gls{r}]{\gls{r} \given \gls{ss}}  = \underset{\gls{ss} \in \{0,1\}}{\text{argmax}} \;  \underset{\gls{siui}}{\mathbb{E}} \{\fpdf[\gls{r}]{\gls{r} \given \gls{ss}, \gls{siui}}\} \nonumber \\
  &=  \underset{\gls{ss} \in \{0,1\}}{\text{argmax}}\mkern-4.5mu \sum\limits_{\gls{siui} \in \mathcal{M} }\mkern-4.5mu \fpdf[\gls{r}]{\gls{r} \mkern-3.5mu\given\mkern-3.5mu \gls{ss}, \gls{siui}} \fpdf[\gls{siui}]{\gls{siui}}\mkern-4.5mu \label{eq:math_section_decision_metric} \\
  &= \mkern-4.5mu\begin{cases}
      1 , \text{if} \,\frac{\sum\limits_{\gls{siui} \in \mathcal{M} } \fpdf[\gls{r}]{\gls{r} \given \gls{ss}=1, \gls{siui}} \fpdf[\gls{siui}]{\gls{siui}}}{\sum\limits_{\gls{siui} \in \mathcal{M} } \fpdf[\gls{r}]{\gls{r} \given \gls{ss}=0, \gls{siui}} \fpdf[\gls{siui}]{\gls{siui}}} \geq 1 \\
       0 , \text{otherwise}
  \end{cases} \nonumber\\
  &= \begin{cases}
      1 , \text{if} \,\frac{\sum\limits_{\gls{siui} \in \mathcal{M} } {(\gls{csmean} + \gls{siui}\gls{cmeaniuiTrans} + \gls{cmeann})}^{\gls{r}} \exp\left(-(\gls{csmean} + \gls{siui}\gls{cmeaniuiTrans} + \gls{cmeann})\right) }{\sum\limits_{\gls{siui} \in \mathcal{M}} {(\gls{siui}\gls{cmeaniuiTrans} + \gls{cmeann})}^{\gls{r}} \exp\left(-(\gls{siui}\gls{cmeaniuiTrans} + \gls{cmeann})\right) } \geq 1 \\
       0 , \text{otherwise}
  \end{cases}\mkern-13.5mu,
  \label{eq:math_section_ML}
\end{align}
where $\fpdf[\gls{r}]{\gls{r} \given \gls{ss}}$ denotes the distribution of $\gls{r}$ in \Equation{eq:system_model_received_molecules} conditioned on \gls{ss}, which is a Poisson distribution with mean $\gls{ss}\gls{csmean}  + \gls{siui}\gls{cmeaniuiTrans} + \gls{cmeann}$ as \gls{cs}, \gls{ciui}, and \gls{cn} are all Poisson distributed, cf. \EquationsList{equation:math_sec:ss_pois}{equation:math_sec:n_pois}. Here, $\fpdf[\gls{siui}]{\gls{siui}}$, $\underset{\gls{siui}}{\mathbb{E}}\{\cdot\}$, and $\mathcal{M} = \{0,1\}^{\gls{nrTrans}-1}$ denote the joint distribution of the transmit symbols of the interferers, the expectation \ac{w.r.t.} $\gls{siui}$, and the set of all possible interference symbol vectors $\gls{siui}$, respectively. The \ac{ML} decision rule in \Equation{eq:math_section_ML} is computationally complex as all possible realizations of the \ac{ILI} have to be taken into account and \Equation{eq:math_section_ML} has to be computed for every received \gls{r}. Note that evaluating \Equation{eq:math_section_ML} is only feasible for a finite number of interferers, cf. discussion in \Section{ssSec:IUIMol}.

In the following, we show that \Equation{eq:math_section_ML} can be equivalently realized by a multi-threshold detector employing a set of pre-calculated threshold values. The threshold values specify the limits of decision regions, where each region corresponds to either $\gls{ss}=1$ or $\gls{ss}=0$. In principle, for complicated probability distributions $\fpdf[\gls{r}]{\gls{r} \given \gls{ss}}$, many regions, and thus many thresholds, are needed. However, we later show in \Corollary{corolarry_threshold} that for special cases the number of thresholds required reduces to a single threshold. Threshold detection is preferable to \Equation{eq:math_section_ML} as the threshold for a given setup can be computed offline and can then be used throughout the transmission. Therefore, the computational cost for online data detection is reduced.
\begin{lem}\label{lemma_multi_threshold}
  The \ac{ML} decision rule given in \Equation{eq:math_section_ML} can be written equivalently as a multi-threshold detection where the finite set of threshold values $T$, with $\lceil \Phi \rceil \in T$ and $ \Phi \in \mathbb{R}^+_0 $, are obtained as the solutions of the following equation
    \scaleAlign
      \begin{align}
          \sum\limits_{\gls{siui} \in \mathcal{M}}\mkern-10.5mu{(\gls{csmean} + \gls{siui}\gls{cmeaniuiTrans} + \gls{cmeann})}^{\Phi} \mkern-4.5mu\exp\mkern-4.5mu\left(-(\gls{csmean} + \gls{siui}\gls{cmeaniuiTrans} + \gls{cmeann})\right)\mkern-4.5mu \nonumber\\
           = \mkern-10.5mu\sum\limits_{\gls{siui} \in \mathcal{M}} \mkern-10.5mu{(\gls{siui}\gls{cmeaniuiTrans} + \gls{cmeann})}^{\Phi} \mkern-4.5mu\exp\mkern-4.5mu\left(-(\gls{siui}\gls{cmeaniuiTrans} + \gls{cmeann})\right) .
          \label{eq:general_Threshold}
      \end{align}
\end{lem}
\begin{IEEEproof}
  The number of obtained threshold values is finite if \gls{nrTrans} is truncated to a finite number according to \Section{ssSec:IUIMol}, i.e., $|T| < \infty$, where $|\cdot|$ denotes the cardinality of a set.
  $|T|$ is upper bounded by the cardinality of $\mathcal{M}$, i.e., $|\mathcal{M}| = 2^{\gls{nrTrans}-1}$, which is the number of possible maxima of the distributions in \Equation{eq:general_Threshold}.
  Due to the symmetry in the proposed \ac{TX}/\ac{RX} grids, some of the interferers have identical statistical impact on \gls{RX0}. Therefore, $|T| < 2^{\gls{nrTrans}-1}$ follows in general.
\end{IEEEproof}
\begin{corollary}\label{corolarry_threshold}
  If $\frac{\gls{csmean}}{\mathbf{1}\gls{cmeaniuiTrans}} = \textrm{SINR}_{\textrm{worst}} > 1$, where $\mathbf{1}$ denotes the \mbox{$ (\gls{nrTrans}-1)$-dimensional} all-ones row vector, the \ac{ML} decision rule given in \Equation{eq:math_section_ML} can be equivalently written as a \textit{single} threshold detection as follows
\scaleAlign
  \begin{align}
    \gls{ssHat} &=
        \begin{cases}
          1 , \; \text{if} \; \gls{r} \geq  \gls{threshold}_{\mathrm{opt}, \gls{nrTrans}-1}\\
          0 , \;\text{otherwise} \;
        \end{cases}\;,
        \label{eq:math_section:threshold}
  \end{align}
  with threshold value $\gls{threshold}_{\mathrm{opt}, \gls{nrTrans}-1}$ obtained as
\scaleAlign
  \begin{align}
    &\gls{threshold}_{\mathrm{opt}, \gls{nrTrans}-1}
    = \text{min} \Big\{ \gls{threshold} \in \mathbb{N} \;|   \nonumber \\
    &\sum_{\gls{siui} \in \mathcal{M}}{(\gls{csmean} + \gls{siui}\gls{cmeaniuiTrans} + \gls{cmeann})}^{\gls{threshold}}  \exp\left(-(\gls{csmean} + \gls{siui}\gls{cmeaniuiTrans} + \gls{cmeann})\right)  \nonumber \\[-0.3cm]
    & \hspace{0.7cm}  \geq \sum_{\gls{siui} \in \mathcal{M}} {(\gls{siui}\gls{cmeaniuiTrans} + \gls{cmeann})}^{\gls{threshold}}  \exp\left(-(\gls{siui}\gls{cmeaniuiTrans} + \gls{cmeann})\right)  \Big\} \;.
    \label{eq:math_section_threshold_definition}
  \end{align}
\end{corollary}
\begin{IEEEproof}
  Please refer to \Appendix{section:appendix:ThresholdProof}.
\end{IEEEproof}
Here, $\textrm{SINR}_{\textrm{worst}} > 1$ is a sufficient, but not a necessary condition. This means that if the condition does not hold, i.e., $\textrm{SINR}_{\textrm{worst}} \leq 1$, detection based on a single threshold may still be equivalent to the \ac{ML} decision rule in \Equation{eq:math_section_ML}, cf. Lemma~\ref{lemma_multi_threshold}, but we can not guarantee this mathematically.
However, for all numerical results shown in this work, we have observed equivalence between \Equation{eq:general_Threshold} and \eqref{eq:math_section_threshold_definition}, i.e., we observed $|T| = 1$ for all considered system settings.
A simplified but suboptimal decision threshold is obtained by neglecting the randomness of the \ac{ILI}. In this case, the threshold value is obtained using the mean of \gls{ciui} instead of the actual distribution of the \ac{ILI}. Therefore, the decision rule in \Equation{eq:math_section_ML} simplifies to
\scaleAlign
\begin{align}
  \gls{ssHat}_{,\mathrm{sub}} \mkern-5.5mu=\mkern-5.5mu \begin{cases}\mkern-5.5mu
      1 , \; \text{if} \; \frac{{(\gls{csmean} + 0.5 \;\mathbf{1}\,\gls{cmeaniuiTrans} + \gls{cmeann})}^{\gls{r}} \exp\left(-(\gls{csmean} + 0.5 \;\mathbf{1}\,\gls{cmeaniuiTrans} + \gls{cmeann})\right)}{ {(0.5 \;\mathbf{1}\,\gls{cmeaniuiTrans} + \gls{cmeann})}^{\gls{r}} \exp\left(-(0.5 \;\mathbf{1}\,\gls{cmeaniuiTrans} + \gls{cmeann})\right)} \mkern-5.5mu\geq\mkern-5.5mu 1 \\
      \mkern-5.5mu 0 , \;\text{otherwise}
  \end{cases}.
  \label{ml_simple}
\end{align}
Similar to \Equation{eq:math_section:threshold}, \Equation{ml_simple} can be equivalently written as a threshold detection scheme with threshold \cite[Eq. (5)]{Jamali2018NonCoherentDF}
\scaleAlign
\begin{align}
  \gls{threshold}_{\mathrm{sub}, \gls{nrTrans}-1} = \frac{\gls{csmean}}{\ln\left(1+\frac{\gls{csmean}}{ 0.5 \;\mathbf{1} \,\gls{cmeaniuiTrans} + \gls{cmeann}}\right)} \;.
  \label{eq:math_section:simpleThres}
\end{align}
We note that the suboptimal threshold in \Equation{eq:math_section:simpleThres} accurately approximates \Equation{eq:math_section_threshold_definition} for the extreme case of a large number of non-distinguishable interference signals at \gls{RX0}, which corresponds to a scenario of densely packed transmission links. In such a scenario, it is sufficient to consider the \textit{average} \ac{ILI}. We compare the performance of the detector based on the decision thresholds derived in \Equation{eq:math_section_threshold_definition} and \Equation{eq:math_section:simpleThres} in \Section{sec:threshold_value_derivation}.

\scaleSubsection
\subsection{BER for one \ac{TX}-\ac{RX} Pair}
\label{Bit_error_rate_derivation}
\scaleSubsectionBelow
The \ac{BER} for one \ac{TX}-\ac{RX} link can be expressed as follows
\scaleAlign
\begin{align}
    \gls{errorProb}&= \underset{\gls{ss}}{\mathbb{E}}\{\underset{\gls{siui}}{\mathbb{E}}\{ \gls{errorProb}(\gls{ssHat} \given \gls{siui}, \gls{ss}) \}\} \nonumber \\
    &= \sum_{\gls{ss}} \sum_{\gls{siui}} \gls{errorProb}(\gls{ssHat} \given \gls{siui}, \gls{ss}) \fpdf[\gls{siui}]{\gls{siui}} \fpdf[\gls{ss}]{\gls{ss}} \;,
\end{align}
where $\fpdf[\gls{ss}]{\gls{ss}}$ and $\gls{errorProb}(\gls{ssHat} \given \gls{siui}, \gls{ss})$ denote the distribution of \gls{ss} and the error probability conditioned on both the transmitted symbol $\gls{ss}$ and interference vector $\gls{siui}$, respectively. In the following, we derive the \ac{BER} for the detector based on the decision thresholds proposed in \Equation{eq:math_section_threshold_definition} and \Equation{eq:math_section:simpleThres}.
\begin{prop}
  For the proposed threshold detectors, \gls{errorProb} can be expressed as follows
  \begin{numcases}{\gls{errorProb} = }
        \frac{1}{2} \Big( \underbrace{\mathcal{Q}(\gls{threshold}',\gls{csmean} + \gls{cmeann})}_{q} + \underbrace{(1 -  \mathcal{Q}(\gls{threshold}', \gls{cmeann}))}_{p}\Big) , \; \text{if} \; \gls{nrTrans} = 1 \hspace{-2.2cm} \nonumber \\[-0.5cm]\label{eq:performance:error_prob_IUI_free} \\
        \nonumber \\[-0.5cm]
       \frac{1}{2} \Big(\mkern-4.5mu\underbrace{\frac{1}{2^{\gls{nrTrans}-1}} \mkern-10.5mu \sum\limits_{\gls{siui} \in \mathcal{M}} \mkern-10.5mu\mathcal{Q}(\gls{threshold}',\gls{csmean} \mkern-2.5mu+\mkern-2.5mu \gls{siui}\gls{cmeaniuiTrans} \mkern-2.5mu+\mkern-2.5mu \gls{cmeann})}_{q} \mkern-2.5mu\nonumber \\+\mkern-2.5mu \underbrace{\frac{1}{2^{\gls{nrTrans}-1}} \mkern-10.5mu \sum\limits_{\gls{siui} \in \mathcal{M}}\mkern-8.5mu   (1 -  \mathcal{Q}(\gls{threshold}', \gls{siui}\gls{cmeaniuiTrans} \mkern-2.5mu+\mkern-2.5mu \gls{cmeann}))}_{p}\mkern-4.5mu\Big), \;\text{else}  \hspace{-0.2cm}\label{eq:performance:error_prob}
  \end{numcases}\label{eq:both}
  where $\gls{threshold}' = \gls{threshold}_{\mathrm{opt}, \gls{nrTrans}-1}$ and $\gls{threshold}' = \gls{threshold}_{\mathrm{sub}, \gls{nrTrans}-1}$ when the decision thresholds in \Equation{eq:math_section_threshold_definition} and \Equation{eq:math_section:simpleThres} are used, respectively. Here, $\mathcal{Q}(a,b) $ denotes the regularized Gamma function.
\end{prop}

\begin{IEEEproof}
  Please refer to \Appendix{section:appendix:BERProof}.
\end{IEEEproof}
The obtained error probabilities $p$ and $q$ can now be used to calculate the link rate and, subsequently, the \ac{ARE} and \ac{ARTE} in \Equation{eq:performacne_user_rate}, \Equation{eq:performance:effRate}, and \Equation{eq:performance:achievableRate}, respectively. Hence, the proposed multi-link system can be evaluated in terms of the performance metrics introduced in \Section{Performance_metrics}.

\scaleSection
\section{Performance Evaluation}\label{sec:evaluation}
\scaleSectionBelow
In this section, we first specify the simulation setup. Then, we evaluate the analytical expression for the \ac{CIR} in \Equation{eq:math_ir_general} and compare it to results from \ac{PBS}. Subsequently, the analytical expressions for the \ac{BER} in \Equation{eq:performance:error_prob}, the \ac{ARE} in \Equation{eq:performance:effRate}, and the \ac{ARTE} in \Equation{eq:performance:achievableRate} are evaluated.
%
%
\scaleSubsection
\subsection{Simulation Parameters}\label{section:evaluation:PBSAndParam}
\scaleSubsectionBelow

\begin{table*}[!tbp]
  \caption{Default Parameter Values}\label{Table:Parameter}
  \begin{center}
  {\def\arraystretch{1.4}\tabcolsep=8pt
  \begin{tabular}{|| p{.07\textwidth}  |  p{.4\textwidth}  | p{.3\textwidth} ||}
    \hline
    Variable & Definition & Value \\ \hline \hline
    $\gls{r0minHex}$ & Cell centers distance in hexagonal grid  & $4\, \times 10^{-6}\,\si{\metre}$ \\ \hline
    $b$ &  Cell centers distance in square grid  & $ \gls{r0minHex}  \, \sqrt{\frac{\sqrt{3}}{2}} $ \\ \hline
    $\gls{SRX}$ & Receiver radius \scriptsize{(neighboring \ac{RXs} touch each other)}  & $ \frac{c}{2}$ \scriptsize{\&} $ \frac{b}{2}$ \scriptsize{in hexagonal \& square grid, respectively} \\ \hline
    $\gls{LRX}$ & Receiver length   & $4 \, \times 10^{-6}\,\si{\metre}$ \\ \hline
    $d$ &  Distance between \ac{TX} and \ac{RX} plane  & $1 \,\times 10^{-5}\,\si{\metre}$\\ \hline
    $\gls{flow}$ &  Flow velocity   & $6 \, \times 10^{-5}\,\si{\metre \per \second}$ \\ \hline
    $\gls{diffusion}$ &  Diffusion coefficient   & $6 \, \times 10^{-11} \,\si{\square \metre \per \second}$ \\ \hline
    $\gls{nrMol}$ &  Number of released molecules   & $100$ \\ \hline
    $C_{\mathrm{noise}}$ &  Background noise concentration   & $ 0\,\si{\per \cubic \metre}$ \\ \hline
    $\Delta t$ & Time step \ac{PBS} \& Monte Carlo simulation & $10^{-4}\,\si{\second}$ \\ \hline
    $T_{\mathrm{sim}}$ & Simulation time & $1\,\si{\second}$ \\  \hline
    $\eta$ & Maximum truncation error factor in \Equation{eq:math_ir_general} & $10^{-6}$ \\ \hline
  \end{tabular}}
\end{center}
\end{table*}

The default values of the channel parameters are given in \Table{Table:Parameter} and are used if not specified otherwise. Micro-scale parameter values are chosen similar to other MC works \cite{Koo2016MoMIMO,Gursoy2019IndMod}. We emphasize that, by default, \gls{SRX} is chosen to be as large as possible, such that the \ac{RX} touches the cell boundary, i.e., $\gls{SRX} = \frac{\gls{r0minHex}}{2}$ and $\gls{SRX} = \frac{b}{2}$ for the hexagonal and the square grid, respectively. Furthermore, we adopt the hexagonal grid as the default structure.
For the evaluation of the \ac{CIR}, the infinite sum in \Equation{eq:math_ir_general} was truncated to $21$ terms\footnote{The corresponding mismatch error is bounded by factor $\eta$ given in \Table{Table:Parameter}, cf. \Appendix{section:appendix:kmax_truncation}.}, i.e., $k_{\mathrm{max}} = 20$. Furthermore, the threshold value $\gls{threshold}_{\mathrm{opt}, \gls{nrTrans}-1}$ in \Equation{eq:math_section_threshold_definition}, the suboptimal threshold value $\gls{threshold}_{\mathrm{sub}, \gls{nrTrans}-1}$ in \Equation{eq:math_section:simpleThres}, and the \ac{BER} in \Equation{eq:performance:error_prob} were computed for a truncated number of interferers. The numbers of interferers correspond to the first three rings in the respective grid, cf. \Figures{graphic:System_model:Grid_hex}{graphic:System_model:Grid_quad}, i.e., $\gls{nrTrans}-1 = 36$ and $\gls{nrTrans}-1 = 24$ for the hexagonal and square grids, respectively.
\scaleSubsubsection
\subsubsection{Monte Carlo Simulation}
\scaleSubsubsectionBelow
To validate the analytical expressions for the \ac{BER} and \ac{ARE}, and to verify that limiting the number of interferers for numerical evaluation does not affect the accuracy of our analysis, we used Monte Carlo simulation. For Monte Carlo simulation, we considered $\gls{nrTrans}-1 = 1260$ and $\gls{nrTrans}-1 = 840$ interferers for the hexagonal and square grids, respectively. First, we randomly generated up to $I = 5\, \times \,10^5$ realizations of possible transmit symbol vectors $\boldsymbol{s}_{\textnormal{all}} = [\gls{ss}, \gls{siui}]$. Then, we generated the number of molecules observed at \gls{RX0} based on the channel model in \Equation{eq:system_model_received_molecules}, where the random variables $\gls{cs}$ and $\gls{ciui}$ were modeled as Poisson distributed according to \Equations{equation:math_sec:ss_pois}{equation:math_sec:siui_pois}. Parameters $\gls{csmean}$ and $\gls{cmeaniui}$ were obtained from the \ac{CIR} in \Equation{eq:math_ir_general}. For \Section{section:evaulation:ISI}, the additional \ac{ISI} part $c_{\mathrm{ISI}}$ was modeled as Poisson distributed as proposed in \cite[Eq. (1)]{cao2020fractionally}. In particular, here the $c_{\mathrm{ISI}}$ takes into account the interfering symbols from both \gls{TX0} and the interfering transmission links. Next, \Equation{eq:performance:error_prob} was numerically evaluated to determine the \ac{BER} for all possible threshold values up to $100$, i.e., $\gls{threshold}\leq 100$, and the lowest \ac{BER} together with the corresponding threshold value and error probabilities $p$ and $q$ were selected.
\scaleSubsubsection
\subsubsection{Particle-Based Simulation}
\scaleSubsubsectionBelow
To verify the accuracy of the analytical expressions for the \ac{CIR} in \Equation{eq:math_ir_general}, \ac{3D} stochastic \ac{PBS}s were carried out \cite{andrews2009accurate}. In the \ac{PBS}s, the signaling molecules propagate in a three-dimensional open space environment and are affected by both uniform flow and diffusion. The results from \ac{PBS} were averaged over 3000 realizations.
%
%
\scaleSubsection
\subsection{Evaluation of \ac{CIR} and \ac{ILI} Truncation} \label{sec:system_model_verification}
\scaleSubsectionBelow
\scaleSubsubsection
\subsubsection{Verification of \ac{CIR}} \label{sec:CIR_verification}
\scaleSubsubsectionBelow

\begin{figure*}[!tbp]
    \begin{minipage}[t]{0.485\textwidth}
      \centering
        \includegraphics[trim=0cm 0cm 1cm 1cm, width=0.95\textwidth]{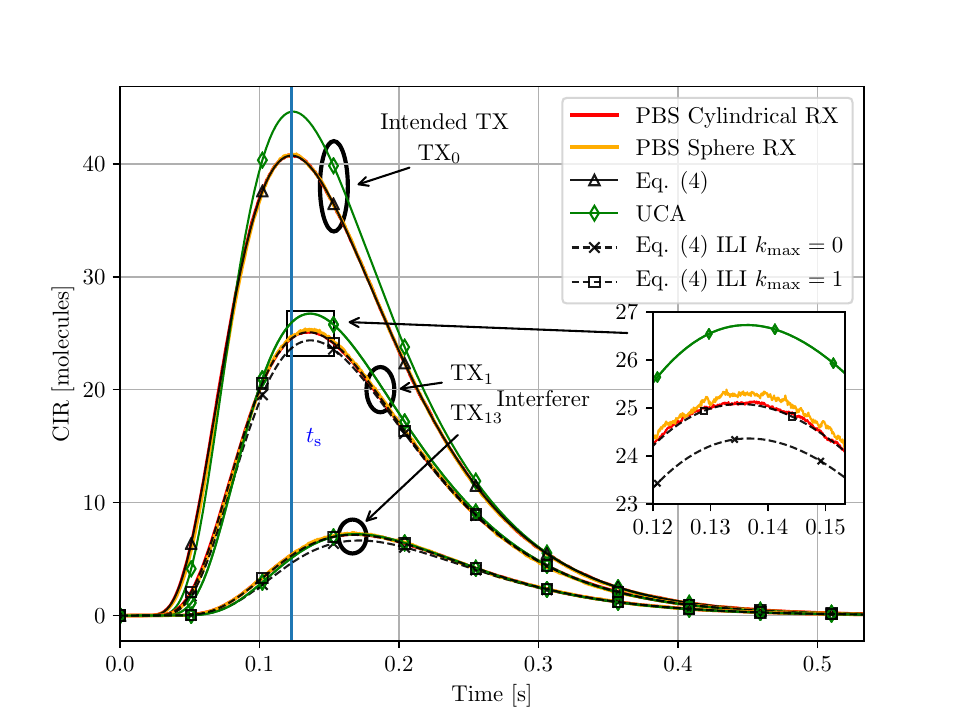}
        \caption{\acs{CIR} from \eqref{eq:math_ir_general} for $\acs{TX}_0$ (solid black) and for $\acs{TX}_1$ and $\acs{TX}_{13}$ (dashed black) realized by a truncation of the sum in \eqref{eq:math_ir_general} to $k_{\mathrm{max}}=0$ and $k_{\mathrm{max}}=1$, UCA (green), and \acs{PBS} with spherical (yellow) and cylindrical (red) RX, respectively, with sampling time $\gls{samplingTime}$ (blue). The diffusion coefficient is set to $D = 6 \, \times 10^{-11} \,\si{\square \metre \per \second}$.}\label{graphic:evaluation:CIR_D_001}
    \end{minipage}
    \hfill
    \begin{minipage}[t]{0.485\textwidth}
      \centering
        \includegraphics[trim=0cm 0cm 1cm 1cm, width=0.95\textwidth]{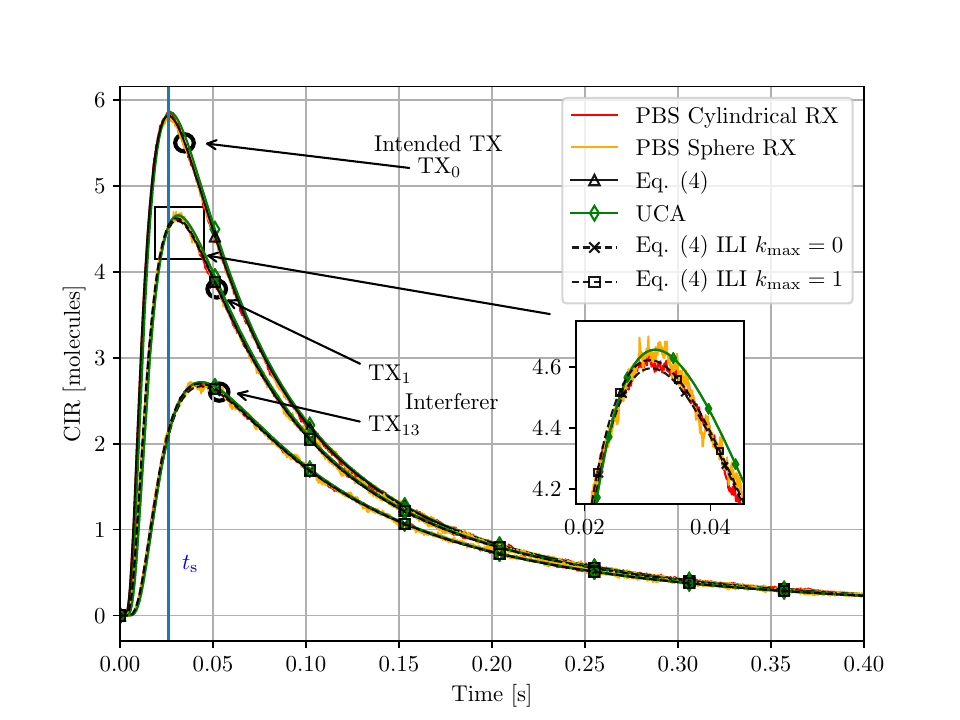}
        \caption{\acs{CIR} from \eqref{eq:math_ir_general} for $\acs{TX}_0$ (solid black) and for $\acs{TX}_1$ and $\acs{TX}_{13}$ (dashed black) realized by a truncation of the sum in \eqref{eq:math_ir_general} to $k_{\mathrm{max}}=0$ and $k_{\mathrm{max}}=1$, UCA (green), and \acs{PBS} with spherical (yellow) and cylindrical (red) RX, respectively, with sampling time $\gls{samplingTime}$ (blue). The diffusion coefficient is set to $D =6 \, \times 10^{-10} \,\si{\square \metre \per \second}$.}\label{graphic:evaluation:CIR_D_01}
    \end{minipage}
\end{figure*}

\Figures{graphic:evaluation:CIR_D_001}{graphic:evaluation:CIR_D_01} show the \ac{CIR} obtained from the proposed analytical expression \Equation{eq:math_ir_general} (black) and \ac{PBS} for cylindrical RXs (red) as ground truth. Additionally, \ac{PBS} results for spherical RXs (yellow) are shown as reference. The radius of the spherical RXs, $S_{\mathrm{RX, Sphere}}$, is chosen such that the volume sizes of the cylindrical and the spherical RXs are identical, which leads to $S_{\mathrm{RX, Sphere}}= 2.289\,\times 10^{-6} \,\si{\meter}$. Here, the CIR of $\acs{TX}_0$ can be directly obtained from the analytical expression in \eqref{eq:math_ir_general}\footnote{Note that the CIR for \gls{TX0} is independent of $k_{\mathrm{max}}$ as the sum in \eqref{eq:math_ir_general} reduces for \gls{TX0} to the term $\gamma(1,\frac{\gls{SRX2}}{4 \gls{diffusion} t})$ as $r_{0}^2 = 0$.} (black solid line), while the CIRs of $\acs{TX}_1$ and $\acs{TX}_{13}$ (black dashed lines) are obtained by an approximation of \eqref{eq:math_ir_general} via a truncation of the sum to $1$ and $2$ terms, i.e., $k_{\mathrm{max}}=0$ and $k_{\mathrm{max}}=1$. Furthermore, we show results for a \ac{CIR} obtained based on the \ac{UCA} at the \ac{RX} (green) as a reference, where the molecule concentration at the center of the \gls{RX0} is scaled with the volume of the \ac{RX} instead of integrating the concentration over the \ac{RX} volume. The \ac{UCA} based \ac{CIR} is obtained as $\mathrm{CIR}_{\mathrm{UCA},i} =  C(x_{\mathrm{RX}_0}, x_{\gls{curTrans}}, y_{\mathrm{RX}_0}, y_{\gls{curTrans}}, z_{\mathrm{RX}}, z_{\gls{curTrans}}, t, t_{0})  \gls{VRX} $  \cite{Jamali2019ChannelMF}. The sampling time $\gls{samplingTime}$ is also depicted (blue). The verification of the \ac{CIR} is exemplarily done for $\ac{TX}_0$, $\ac{TX}_1$, and $\ac{TX}_{13}$ for the default hexagonal grid with cell-center distance $\gls{r0minHex} = 4\, \times 10^{-6}\,\si{\metre}$, see \Figure{graphic:System_model:Grid_hex}, and the two different diffusion coefficients $D = 6 \, \times 10^{-11} \,\si{\square \metre \per \second}$, cf. \Figure{graphic:evaluation:CIR_D_001}, and $D =6 \, \times 10^{-10} \,\si{\square \metre \per \second}$, cf. \Figure{graphic:evaluation:CIR_D_01}.

We first concentrate on the \ac{CIR}s for default diffusion coefficient $D = 6 \, \times 10^{-11} \,\si{\square \metre \per \second}$. From \Figure{graphic:evaluation:CIR_D_001}, we observe that for the desired transmitter \gls{TX0}, the proposed \ac{CIR} is in excellent agreement with the PBS results. For $\ac{TX}_1$ and $\ac{TX}_{13}$, \Equation{eq:math_ir_general} deviates from the \ac{PBS} result for $k_{\mathrm{max}}=0$, but both match for $k_{\mathrm{max}}=1$, i.e., for truncating the infinite sum to two terms\footnote{ We note that only for these results, which are based on severe truncation to $k_{\mathrm{max}}=0$ and $k_{\mathrm{max}}=1$, the truncation error factor is larger than $\eta = 10^{-6}$, while for all other results, which are based on $k_{\mathrm{max}}=20$, $\eta = 10^{-6}$ holds.}. The \ac{UCA} based \ac{CIR} deviates from the \ac{PBS} result, especially for $\ac{TX}_0$ and $\ac{TX}_1$. For $\ac{TX}_{13}$, the \ac{UCA} based \ac{CIR} approximates \Equation{eq:math_ir_general} well. In general, the expression for $\mathrm{CIR}_{\mathrm{UCA},i}$ is more compact than \Equation{eq:math_ir_general}, but less accurate.
These observations are also valid for the \ac{CIR}s for $D =6 \, \times 10^{-10} \,\si{\square \metre \per \second}$, which are shown in \Figure{graphic:evaluation:CIR_D_01}. Furthermore, by comparing \Figures{graphic:evaluation:CIR_D_001}{graphic:evaluation:CIR_D_01}, we observe that the deviation between $\mathrm{CIR}_{\mathrm{UCA},i}$ and the \ac{PBS} result is small for larger diffusion coefficients. For small \gls{diffusion}, the spread of the spatial distribution of the released molecules does not significantly change during molecule transport which is mainly determined by flow. As we assume a point release at the \ac{TX}, i.e., a spatially concentrated molecule distribution, the gradient of the molecule concentration within the receiver volume is large for small diffusion coefficients and contradicts the assumptions underlying the \ac{UCA} \cite{Jamali2019ChannelMF}. We observe that the peak \ac{CIR} value for \gls{TX0} decreases when \gls{diffusion} is increased, as in this case, molecules spread out more during propagation.
Finally, we conclude that the derived expression for the \ac{CIR} in \eqref{eq:math_ir_general} is accurate and in perfect agreement with the ground truth results from PBSs shown in red color. Additionally, we observe a large similarity between the results obtained for cylindrical and spherical RXs, respectively. From this we conclude that our results on the exploitation of the spatial resource, shown in the following, generalize to systems with spherical RXs. Moreover, to get accurate results,  visual examination suggests that it may be sufficient to take into account only two terms in the infinite sum in \eqref{eq:math_ir_general} for the parameters considered in \Figures{graphic:evaluation:CIR_D_001}{graphic:evaluation:CIR_D_01}. Nevertheless, we use $k_{\mathrm{max}}=20$ for the analysis in \Section{eval:BER_ARE_Results}, which \textit{mathematically} guarantees accuracy, cf. \Appendix{section:appendix:kmax_truncation}.
%
%
\scaleSubsubsection
\subsubsection{Impact of the Number of Interferers}\label{section:evaluation:BER_over_threshold}
\scaleSubsubsectionBelow

\begin{figure*}[!tbp]
    \centering
    \begin{minipage}[t]{0.485\textwidth}
        \centering
        \includegraphics[width=0.95\textwidth]{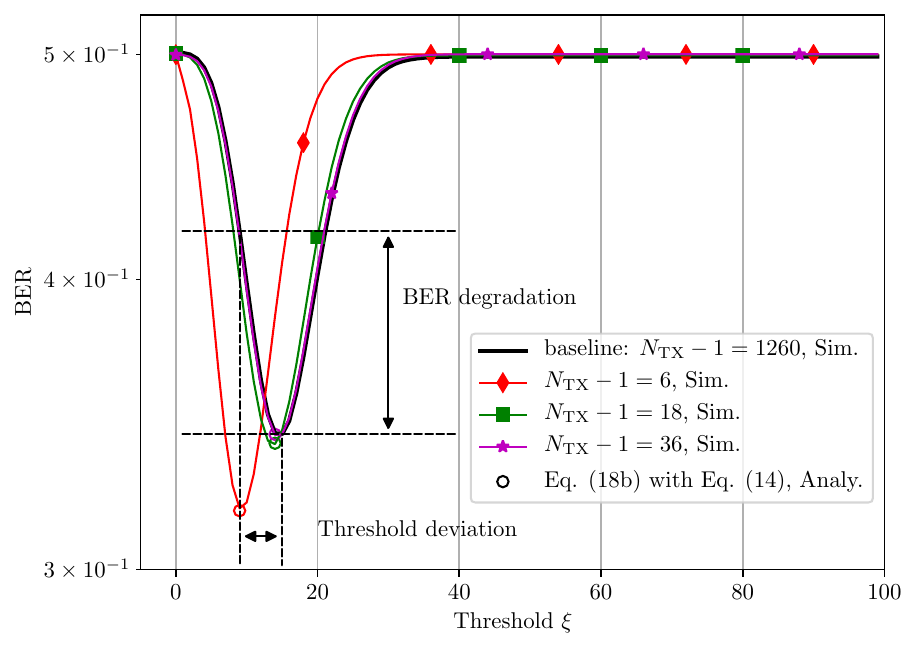}
        \caption{BER vs. threshold \gls{threshold} for different numbers of considered interferers $\gls{nrTrans}-1$ for default cell center distance $\gls{r0minHex} = 4\, \times 10^{-6}\,\si{\metre}$. The result shown as black solid line is referred to as the baseline.}\label{graphic:evaluation:BER_over_threshold_different_rings_difference}
    \end{minipage}
    \hfill
    \begin{minipage}[t]{0.485\textwidth}
        \centering
        \includegraphics[width=0.915\textwidth]{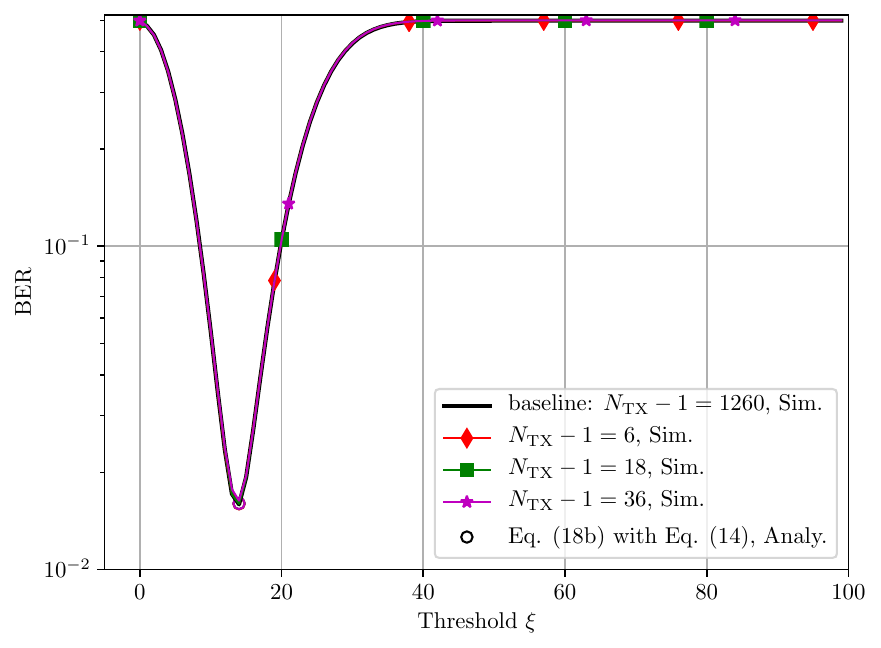}
        \caption{BER vs. threshold \gls{threshold} for different numbers of considered interferers $\gls{nrTrans}-1$ for cell center distance $\gls{r0minHex} = 1\, \times 10^{-5}\,\si{\metre}$. The result shown as black solid line is referred to as the baseline.}\label{graphic:evaluation:BER_over_threshold_different_rings_no_difference}
    \end{minipage}
\end{figure*}

In \Figures{graphic:evaluation:BER_over_threshold_different_rings_difference}{graphic:evaluation:BER_over_threshold_different_rings_no_difference}, the \ac{BER} in \Equation{eq:performance:error_prob} is shown as a function of threshold value \gls{threshold} for different \gls{nrTrans}, i.e., $\gls{nrTrans}-1 = 6$, $\gls{nrTrans}-1 = 18$, $\gls{nrTrans}-1 = 36$ (dashed colored lines), and $\gls{nrTrans}-1 = 1260$ (black color). The latter case is referred to as the baseline in the following. All results shown in \Figures{graphic:evaluation:BER_over_threshold_different_rings_difference}{graphic:evaluation:BER_over_threshold_different_rings_no_difference} were obtained by Monte Carlo simulation. The impact of the truncation of the number of interferers on the accuracy of detection is exemplarily shown for two cell center distances, i.e., $\gls{r0minHex} = 4\, \times 10^{-6}\,\si{\metre}$ and $\gls{r0minHex} = 1\, \times 10^{-5}\,\si{\metre}$. Note that the results obtained via Monte Carlo simulation perfectly match the \ac{ML} detector results obtained analytically by threshold $\gls{threshold}_{\mathrm{opt}, \gls{nrTrans}-1}$ and the \ac{BER} according to \Equation{eq:math_section_threshold_definition} and \Equation{eq:performance:error_prob} (circular marker), respectively.

From \Figure{graphic:evaluation:BER_over_threshold_different_rings_difference}, we observe that, as predicted in \Corollary{corolarry_threshold}, there exists \textit{one} optimal threshold value $\gls{threshold}_{\mathrm{opt}, \gls{nrTrans}-1}$ that minimizes the \ac{BER} for a given number of interfering transmitters $\gls{nrTrans}-1$, although condition $\textrm{SINR}_{\textrm{worst}} > 1$, cf. \Equation{eq:app:sinr_condition} in \Appendix{section:appendix:ThresholdProof}, is not satisfied, i.e., $\textrm{SINR}_{\textrm{worst}} = \{0.276588, 0.175053, 0.163543, 0.162704\} \leq 1$ for $\gls{nrTrans}-1 = \{6, 18, 36, 1260\}$, respectively.
Decoding the received signal by using a larger or smaller threshold value compared to $\gls{threshold}_{\mathrm{opt}, \gls{nrTrans}-1}$, i.e., $\gls{threshold} > \gls{threshold}_{\mathrm{opt}, \gls{nrTrans}-1}$ or $\gls{threshold} < \gls{threshold}_{\mathrm{opt}, \gls{nrTrans}-1}$, respectively, increases the \ac{BER}. We further observe from \Figure{graphic:evaluation:BER_over_threshold_different_rings_difference} that truncating the number of interferers to a smaller value can result in a lower threshold value, i.e., $\gls{threshold}_{\mathrm{opt},6} \leq \gls{threshold}_{\mathrm{opt},18} \leq \gls{threshold}_{\mathrm{opt},36} \leq \gls{threshold}_{\mathrm{opt},1260}$, and a lower \ac{BER}.
Thus, if the number of interferers considered is not sufficient, a threshold deviation results, i.e., $\gls{threshold}_{\mathrm{opt},1260} - \gls{threshold}_{\mathrm{opt},6} > 0$, which leads to a \ac{BER} degradation\footnote{The setup, which considers six interfering transmitting links, can also be interpreted as a system of small extent. Since grid size is kept constant, due to the reduced ILI, such system results in a smaller BER, as expected, cf. \Section{grid_section}.}. The deviation of the threshold value as well as the subsequent performance loss in terms of the \ac{BER} are depicted in \Figure{graphic:evaluation:BER_over_threshold_different_rings_difference}.
From \Figure{graphic:evaluation:BER_over_threshold_different_rings_difference}, we further observe a close match between the \ac{BER} results for $\gls{nrTrans}-1 = 36$ and the baseline case, while there exists a mismatch between the \ac{BER} results for $\gls{nrTrans}-1 = 18$ and the baseline case. Hence, the \ac{ILI} from more than $18$ additional transmission links notably impairs the link between \gls{TX0} and \gls{RX0}, which leads to the obtained larger BER. We finally conclude that truncating the number of interferers to $\gls{nrTrans}-1 = 36$ is sufficient for $\gls{r0minHex} = 4\, \times 10^{-6}\,\si{\metre}$.

In \Figure{graphic:evaluation:BER_over_threshold_different_rings_no_difference}, the agreement between the baseline case and the result obtained by truncation of the number of interferers to  $\gls{nrTrans}-1 = 6$ is excellent. Thus, for $\gls{r0minHex} = 4\, \times 10^{-6}\,\si{\metre}$, $\gls{nrTrans}-1 = 6$ is sufficient to accurately model the impact of \ac{ILI}. Note that here condition \Equation{eq:app:sinr_condition} is satisfied as $\textrm{SINR}_{\textrm{worst}} = \{1.726913, 1.717973, 1.717973, 1.717973\}$ for $\gls{nrTrans}-1 = \{6, 18, 36, 1260\}$, respectively.

The exact number of interferers required to accurately estimate the optimal threshold value $\gls{threshold}_{\mathrm{opt}, \infty}$ for the system depends on the system parameters and should be chosen carefully. A simple method to obtain the necessary number of interferers is by brute force search, i.e., iteratively increasing \gls{nrTrans} and evaluating the \ac{BER} until the increase results in a non-distinguishable change of the \ac{BER}. In particular, we verify this brute force method by additionally showing performance results for $\gls{nrTrans}-1 = 1260$ interferers for comparison in \FigureList{graphic:evaluation:BER_Noise}{graphic:evaluation:rate_subOptThres}. Hence, as long as the results for $36$ and $1260$ interferers are identical, $36$ interferers are sufficient to characterize the entire \ac{ILI}. Note that the results in \Section{eval:BER_ARE_Results} for $\gls{nrTrans}-1 = 36$ interferers are obtained analytically from \Equation{eq:math_section_threshold_definition}, \Equation{eq:performance:error_prob}, and \Equation{eq:performance:effRate}, while the results for $\gls{nrTrans}-1 = 1260$ interferers are obtained by Monte Carlo simulations, if not specified otherwise. The reason for utilizing Monte Carlo simulation for large numbers of interferers is that for $1260$ interferers, the analytical evaluation of the results becomes infeasible due to the $2^{\gls{nrTrans}-1}$ possible realizations of the interference term, which need to be considered, i.e., $2^{36} \approx 6.9 \times 10^{10}$ vs. $2^{1260} \approx 2 \times 10^{379}$.
Finally, we observe from \Figures{graphic:evaluation:BER_over_threshold_different_rings_difference}{graphic:evaluation:BER_over_threshold_different_rings_no_difference} that the achievable \ac{BER} for $\gls{r0minHex} = 1\, \times 10^{-5}\,\si{\metre}$ is about one order of magnitude smaller than that for $\gls{r0minHex} = 4\, \times 10^{-6}\,\si{\metre}$. This is intuitive as the \ac{ILI} decreases for increasing \gls{r0minHex} and therefore the \ac{BER} decreases.


\scaleSubsection
\subsection{Evaluation of BER, ARE, and ARTE}\label{eval:BER_ARE_Results}
In this section, we evaluate the \ac{BER}, \ac{ARE}, and \ac{ARTE} for various scenarios.

\scaleSubsubsection
\subsubsection{Impact of Background Noise on BER and ARE} \label{section:evaulation:BackgroundNoise}

\begin{figure*}[!tbp]
    \centering
    \begin{minipage}[t]{0.485\textwidth}
        \centering
        \includegraphics[trim=0cm 0cm 1cm 1cm, trim=0cm 0cm 1cm 1cm, width=0.95\textwidth]{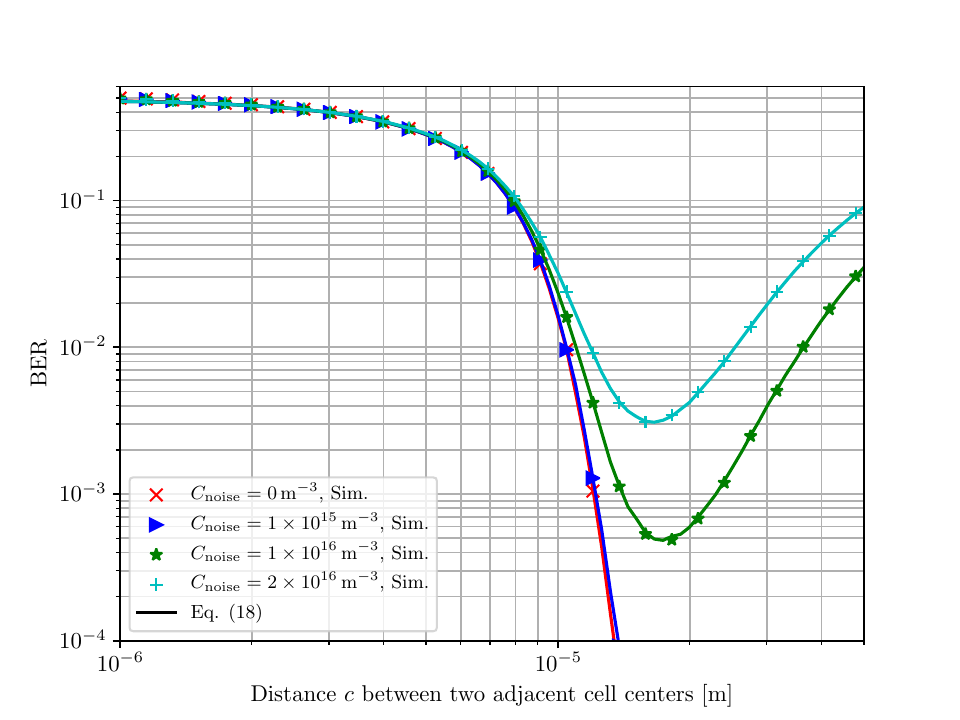}
        \caption{\ac{BER} as a function of cell center distance $\gls{r0minHex}$ for different background noise molecule concentrations $C_{\mathrm{noise}}$. The results from Monte Carlo simulation are depicted by markers.}
        \label{graphic:evaluation:BER_Noise}
    \end{minipage}
    \hfill
    \begin{minipage}[t]{0.485\textwidth}
        \centering
        \includegraphics[trim=0cm 0cm 1cm 1cm, width=0.95\textwidth]{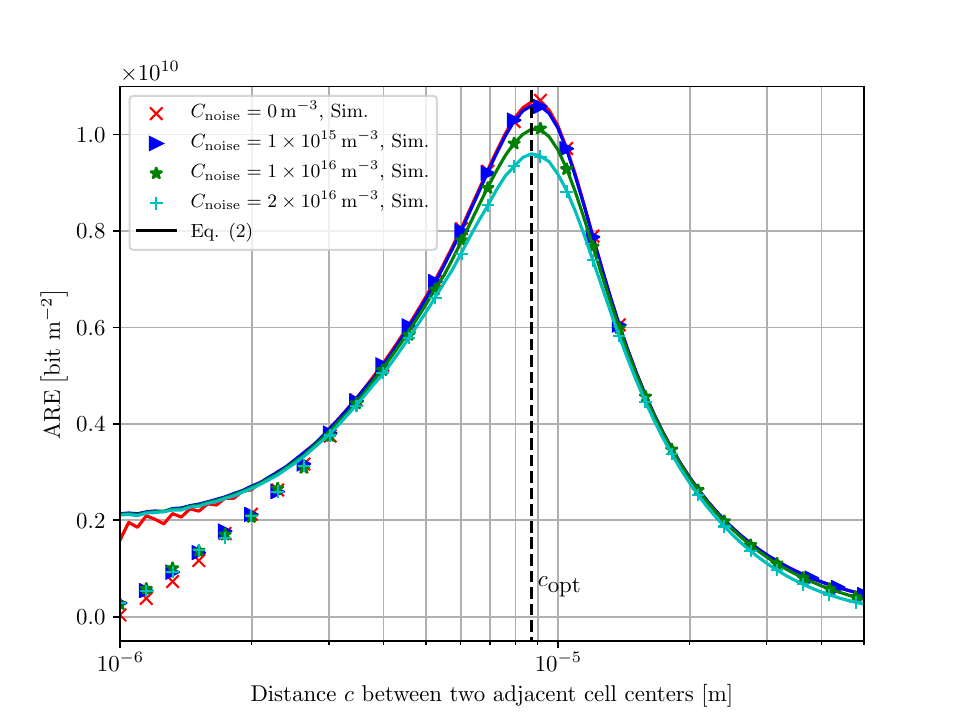}
        \caption{\ac{ARE} as a function of cell center distance $\gls{r0minHex}$ for different background noise molecule concentrations $C_{\mathrm{noise}}$. The results from Monte Carlo simulation are depicted by markers.}
        \label{graphic:evaluation:rate_Noise}
    \end{minipage}

\end{figure*}

\Figures{graphic:evaluation:BER_Noise}{graphic:evaluation:rate_Noise} depict the \ac{BER} and \ac{ARE} as functions of the hexagonal cell center distance \gls{r0minHex} for different background noise molecule concentrations $C_{\mathrm{noise}}$.

We observe from \Figure{graphic:evaluation:BER_Noise} that \ac{BER} approaches $0.5$ for small \gls{r0minHex} and for increasing \gls{r0minHex}, the \ac{BER} decreases in the absence of background noise molecules. However, for $C_{\mathrm{noise}} > 0$, above a certain value of \gls{r0minHex}, the \ac{BER} increases again (see green and cyan curves).
In fact, if a certain cell center distance \gls{r0minHex} with corresponding cell area $A_{\textnormal{cell}}$ is exceeded, a further increase leads to a larger number of received background noise molecules, but not to a larger number of received information carrying molecules, which causes the \ac{BER} to increase.
Furthermore, as expected, the BER increases for increasing number of background noise molecules.

\Figure{graphic:evaluation:rate_Noise} shows that the \ac{ARE} has a unique maximum and we denote the corresponding cell center distance as $\gls{r0minHex}_{\mathrm{opt}}$. We observe that both, decreasing and increasing \gls{r0minHex} compared to $\gls{r0minHex}_{\mathrm{opt}}$, decreases the simulated \ac{ARE} asymptotically to $\ac{ARE} \rightarrow 0$.

For small \gls{r0minHex}, the considered system suffers from excessive \ac{ILI}, i.e., the resulting small $R_{\textnormal{SISO}}$ mainly limits the \ac{ARE} in \Equation{eq:performance:effRate}. However, for large \gls{r0minHex}, the system's usage of the spatial resource is not optimal, as the hexagonal cell area reserved for one \ac{TX}-\ac{RX} link is large, and the small $\gls{spatialRate}$ limits the \ac{ARE} in \Equation{eq:performance:effRate}. From the existence of a maximum \ac{ARE}, we conclude that there exists an optimal \ac{TX}-\ac{RX} link density, which is achieved by \ac{TX} positions with cell center distance $\gls{r0minHex} = \gls{r0minHex}_{\mathrm{opt}}$.
We further observe that the maximum \ac{ARE} decreases for increasing background noise molecule concentration. We note that $\gls{r0minHex}_{\mathrm{opt}}$, and therefore, the optimal \ac{TX}-\ac{RX} link density shows no dependency on the background noise molecule concentration.

Finally, we observe from \Figure{graphic:evaluation:rate_Noise}, that the analytical \ac{ARE} \Equation{eq:performance:effRate} deviates from the Monte Carlo simulation result for small \gls{r0minHex}. The reason for this is that the actual number of interferers needed to accurately approximate the system behavior increases with decreasing \gls{r0minHex}. Hence, for small \gls{r0minHex}, more than $36$ interferers have to be considered in order to properly model the \ac{ILI}.

\scaleSubsubsection
\subsubsection{Hexagonal Grid vs. Square Grid}\label{section:evaulation:Grid}

\begin{figure*}[!tbp]
    \centering
    \begin{minipage}[t]{0.485\textwidth}
        \centering
        \includegraphics[trim=0cm 0cm 1cm 1cm, width=0.95\textwidth]{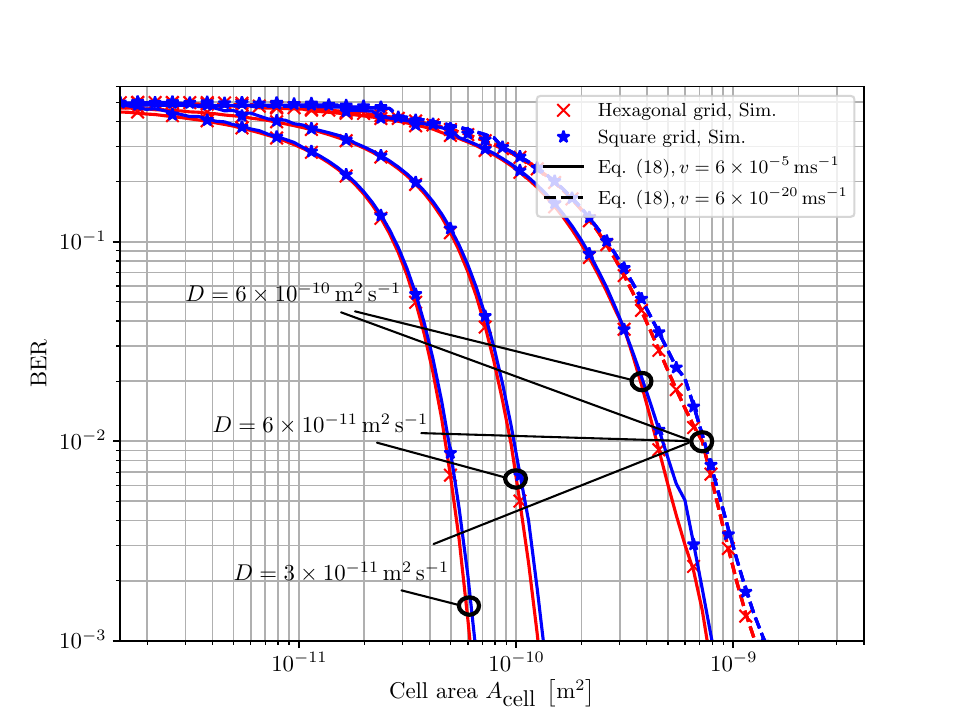}
        \caption{\ac{BER} as a function of cell area $A_{\mathrm{cell}}$ for the two considered grids, hexagonal and square, and different diffusion coefficients \gls{diffusion}. The results from Monte Carlo simulation are depicted by markers.}\label{graphic:evaluation:BER_grid}
    \end{minipage}
    \hfill
    \begin{minipage}[t]{0.485\textwidth}
        \centering
        \includegraphics[trim=0cm 0cm 1cm 1cm, width=0.95\textwidth]{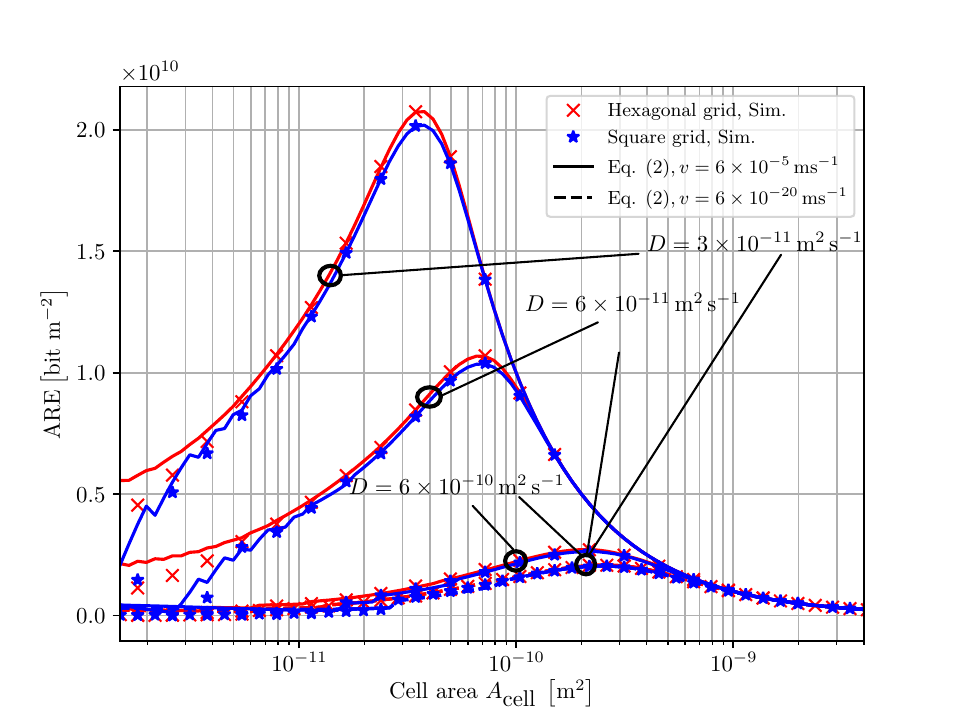}
        \caption{\ac{ARE} as a function of cell area $A_{\mathrm{cell}}$ for the two considered grids, hexagonal and square, and different diffusion coefficients \gls{diffusion}. The results from Monte Carlo simulation are depicted by markers.}\label{graphic:evaluation:rate_grid}
    \end{minipage}

\end{figure*}

In this section, we investigate the impact of the type of grid,  diffusion, and flow velocity on \ac{BER} and \ac{ARE}. As discussed in \Section{sec:model}, we assume equal cell area sizes, i.e., $A_\mathrm{hex} = A_\mathrm{quad} = b^2$. Furthermore, the \ac{RX} radius $\gls{SRX}$ is chosen such that neighboring \ac{RXs} touch, but do not overlap. As a consequence, $\gls{SRX}$ and therefore the \ac{RX} volumes are smaller for the square grid compared to the hexagonal grid, i.e., $\gls{SRX}_{, \mathrm{hex}}> \gls{SRX}_{, \mathrm{quad}}$, and consequently $\gls{VRX}_{, \mathrm{hex}}> \gls{VRX}_{, \mathrm{quad}}$.

In \Figures{graphic:evaluation:BER_grid}{graphic:evaluation:rate_grid}, \ac{BER} and \ac{ARE} are shown as a function of the cell area $A_{\mathrm{cell}}$ for different diffusion coefficients $\gls{diffusion}$ and two different flow velocities for a background noise free scenario. In particular, one flow velocity is very low, $ v = 6 \, \times 10^{-20}\,\si{\metre \per \second}$, i.e., flow is not relevant for the molecule propagation in this case. Therefore, this case is referred to as flow-free scenario in the following. For this scenario, $T_{\mathrm{sim}} = 6 \, \si{\second}$ is used. We observe that the general behaviors of the \ac{BER} and \ac{ARE} are similar as in \FigureList{graphic:evaluation:BER_Noise}{graphic:evaluation:rate_grid}. We denote the cell area that maximizes the \ac{ARE} as $A_{\mathrm{cell}}^{\mathrm{opt}}$.

We first consider the impact of diffusion coefficient $\gls{diffusion}$ with $v =  6\, \times 10^{-5}\,\si{\metre \per \second}$. \Figure{graphic:evaluation:BER_grid} shows that \ac{BER} increases with $\gls{diffusion}$. This is intuitive as \gls{diffusion} characterizes the dispersion, i.e., the spatial molecule spread, which increases with \gls{diffusion} as diffusion gains in importance compared to molecule transport via flow. In particular, for increasing \gls{diffusion}, the number of information molecules observed at the sampling time decreases and simultaneously the number of observed interfering molecules increases at \gls{RX0}. In \Figure{graphic:evaluation:rate_grid}, we observe that the peak of the \ac{ARE} is higher for smaller \gls{diffusion} and is obtained for a smaller cell area $A_{\mathrm{cell}}^{\mathrm{opt}}$.  We further observe that the flow-free scenario yields a larger \ac{BER} and smaller \ac{ARE} compared to the scenario with $v =  6\, \times 10^{-5}\,\si{\metre \per \second}$. Finally, \Figures{graphic:evaluation:BER_grid}{graphic:evaluation:rate_grid} show that for the flow-free scenario varying the diffusion coefficient yields unaltered results. This is intuitive, as for this scenario a change in \gls{diffusion} has no impact on the number of information molecules observed at the peak time, which itself may change.

Next, we consider the effect of the chosen grid structure. We observe from \Figure{graphic:evaluation:BER_grid} that the \ac{BER} for the hexagonal grid is lower than that for the square grid. Furthermore, \Figure{graphic:evaluation:rate_grid} shows a higher maximum \ac{ARE} for the hexagonal grid than for the square grid. Hence, the hexagonal grid is preferable compared to the square grid in terms of \ac{BER} and \ac{ARE}. However, the performance differences between both grid structures are small and diminish for large values of $A_{\mathrm{cell}}$.
Finally, we observe from \Figure{graphic:evaluation:rate_grid} that the analytical results deviate from the Monte Carlo simulations for very small $A_{\mathrm{cell}}$ as in this case not enough interferers are taken into account.

\scaleSubsubsection
\subsubsection{Comparison of Methods to Derive the Threshold Value}\label{sec:threshold_value_derivation}

\begin{figure*}[!tbp]
    \centering
    \begin{minipage}[t]{0.485\textwidth}
        \centering
        \includegraphics[trim=0cm 0cm 1cm 1cm, width=0.95\textwidth]{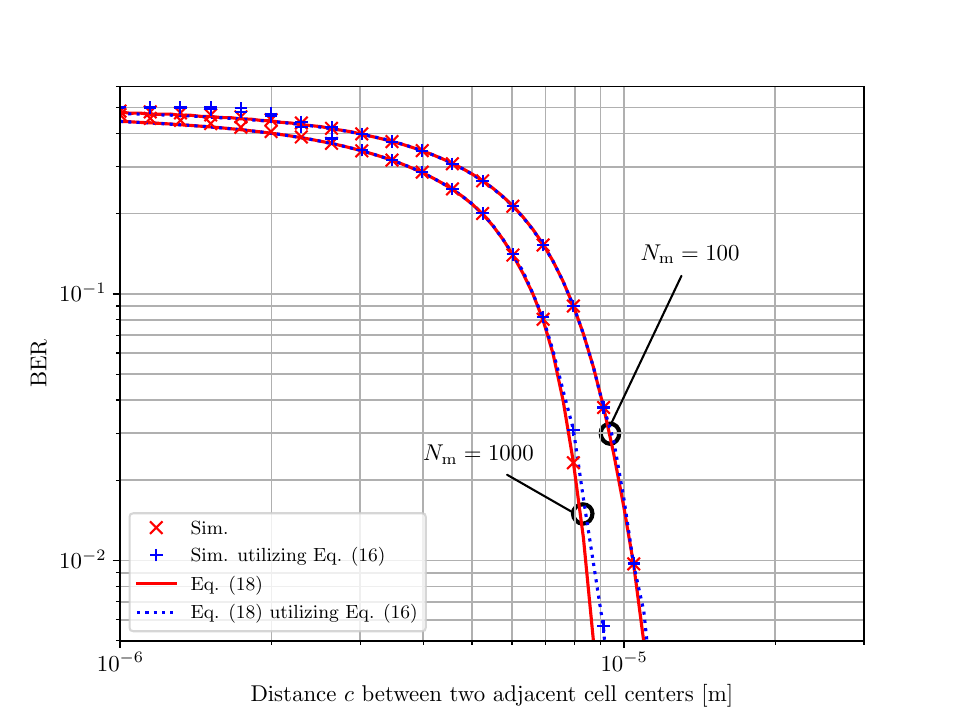}
        \caption{\ac{BER} as a function of cell center distance $\gls{r0minHex}$. The \ac{BER} for the analytical threshold values as defined in \Equation{eq:math_section_threshold_definition} (solid line) and \Equation{eq:math_section:simpleThres} (dashed line) is compared to that for the threshold values obtained by Monte Carlo simulation (markers).}
        \label{graphic:evaluation:BER_subOptThres}
    \end{minipage}
    \hfill
    \begin{minipage}[t]{0.485\textwidth}
        \centering
        \includegraphics[trim=0cm 0cm 1cm 1cm, width=0.95\textwidth]{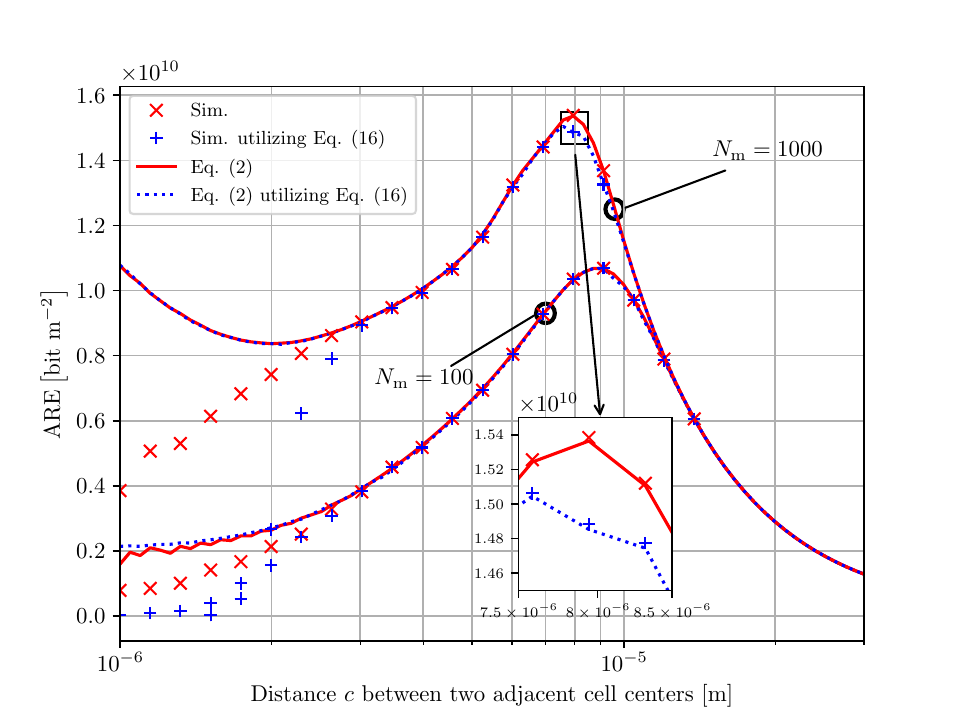}
        \caption{\ac{ARE} as a function of cell center distance $\gls{r0minHex}$. The \ac{ARE} for the analytical threshold values as defined in \Equation{eq:math_section_threshold_definition} (solid line) and \Equation{eq:math_section:simpleThres} (dashed line) is compared to that for the threshold values obtained by Monte Carlo simulation (markers). }
        \label{graphic:evaluation:rate_subOptThres}
    \end{minipage}
\end{figure*}

In this section, we evaluate the loss in performance caused by the suboptimal threshold value $\gls{threshold}_{\mathrm{sub}, \gls{nrTrans}-1}$ in \Equation{eq:math_section:simpleThres}. We further show the impact of the number of released molecules \gls{nrMol} on the \ac{BER} and \ac{ARE}.

\Figures{graphic:evaluation:BER_subOptThres}{graphic:evaluation:rate_subOptThres} show that for small cell center distances \gls{r0minHex}, the \ac{BER} and \ac{ARE} results for $\gls{threshold}_{\mathrm{opt}, \gls{nrTrans}-1}$ and $\gls{threshold}_{\mathrm{sub}, \gls{nrTrans}-1}$ are indistinguishable because of the excessive interference. In particular, for small \gls{r0minHex}, more interferers are relevant and the individual interferers are positioned close to each other and therefore have a similar impact on the information transmission of \gls{TX0}, as the amount of interference is distance-dependent. Hence, for small \gls{r0minHex} an averaging effect occurs, and employing the suboptimal threshold $\gls{threshold}_{\mathrm{sub}, \gls{nrTrans}-1}$, which neglects the higher-order \ac{ILI} statistics, is sufficient. Next, we observe that for increasing \gls{r0minHex}, which corresponds to moderate \ac{ILI}, the usage of  $\gls{threshold}_{\mathrm{sub}, \gls{nrTrans}-1}$ for detection compared to $\gls{threshold}_{\mathrm{opt}, \gls{nrTrans}-1}$ leads to a higher \ac{BER} and consequently to lower \ac{ARE} values. We further observe that the \ac{ARE} degradation increases with increasing \gls{nrMol}.

Next, we focus on the impact of the numbers of released molecules \gls{nrMol} on the \ac{BER} and \ac{ARE}. In \Figure{graphic:evaluation:BER_subOptThres}, we observe that for increasing \gls{nrMol} the \ac{BER} decreases.
\Figure{graphic:evaluation:rate_subOptThres} shows that increasing $\gls{nrMol}$ also increases the peak value of the \ac{ARE}. We observe that $\gls{r0minHex}_{\mathrm{opt}}$ is smaller for larger numbers of released molecules \gls{nrMol}, i.e., the optimal density of the independent transmission links is larger for larger \gls{nrMol}. In particular, \gls{nrMol} linearly scales $ \gls{csmean}$, cf. \Section{ssSec:IM}. Hence, the difference between $\gls{ss} = 1$ and $\gls{ss}=0$ \ac{w.r.t.} the expected number of received molecules, i.e., $\mathbb{E}\{\gls{r}\given \gls{ss} = 1\} - \mathbb{E}\{\gls{r}\given \gls{ss} = 0\}$, increases with increasing \gls{nrMol}, and therefore improves the performance of the threshold based detection. In practical applications, however, there might be an upper limit for \gls{nrMol} due to limited resources.

From this section, we conclude that, as expected, detection with threshold $\gls{threshold}_{\mathrm{opt}, \gls{nrTrans}-1}$ yields a higher performance than with $\gls{threshold}_{\mathrm{sub}, \gls{nrTrans}-1}$. However, the loss in terms of \ac{BER} and \ac{ARE} is small and diminishes for some system settings, such as small numbers of released molecules $\gls{nrMol}$.

\scaleSubsection
\subsection{Impact of the Symbol Duration on BER and ARTE}\label{section:evaulation:ISI}
\scaleSubsectionBelow

\begin{figure*}[!tbp]
    \centering
    \begin{minipage}[t]{0.485\textwidth}
        \centering
        \includegraphics[trim=0cm 0cm 1cm 1cm, width=0.95\textwidth]{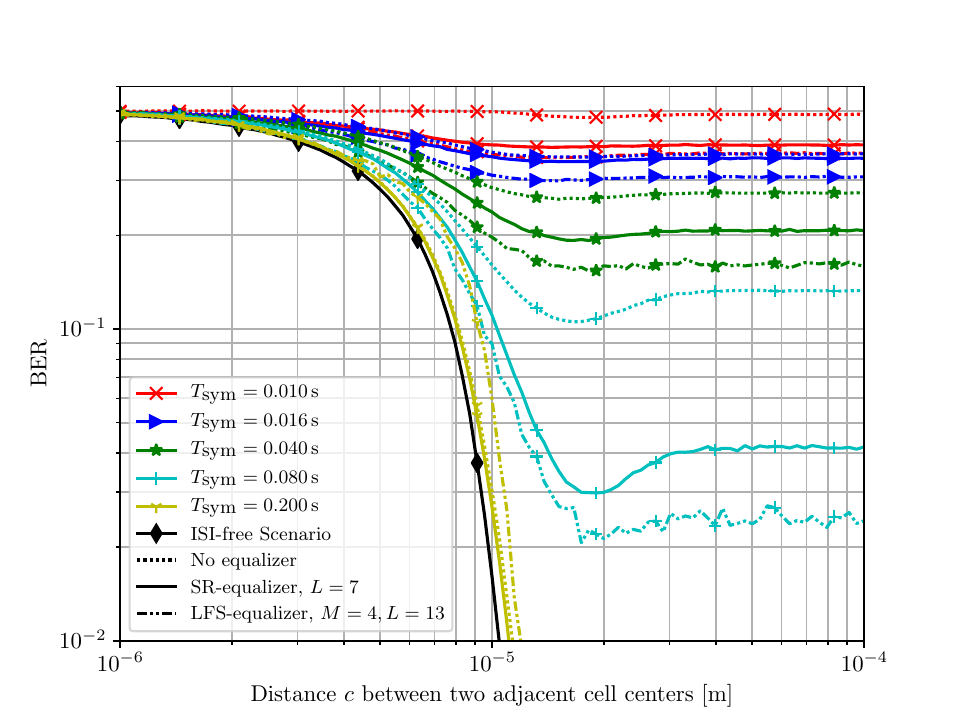}
        \caption{\ac{BER} as a function of cell center distance $\gls{r0minHex}$ for different symbol durations $T_{\mathrm{sym}}$ and different equalization methods. All results are obtained by Monte Carlo simulations.}\label{graphic:evaluation:BER_ISI}
    \end{minipage}
    \hfill
    \begin{minipage}[t]{0.485\textwidth}
        \centering
        \includegraphics[trim=0cm 0cm 1cm 1cm, width=0.95\textwidth]{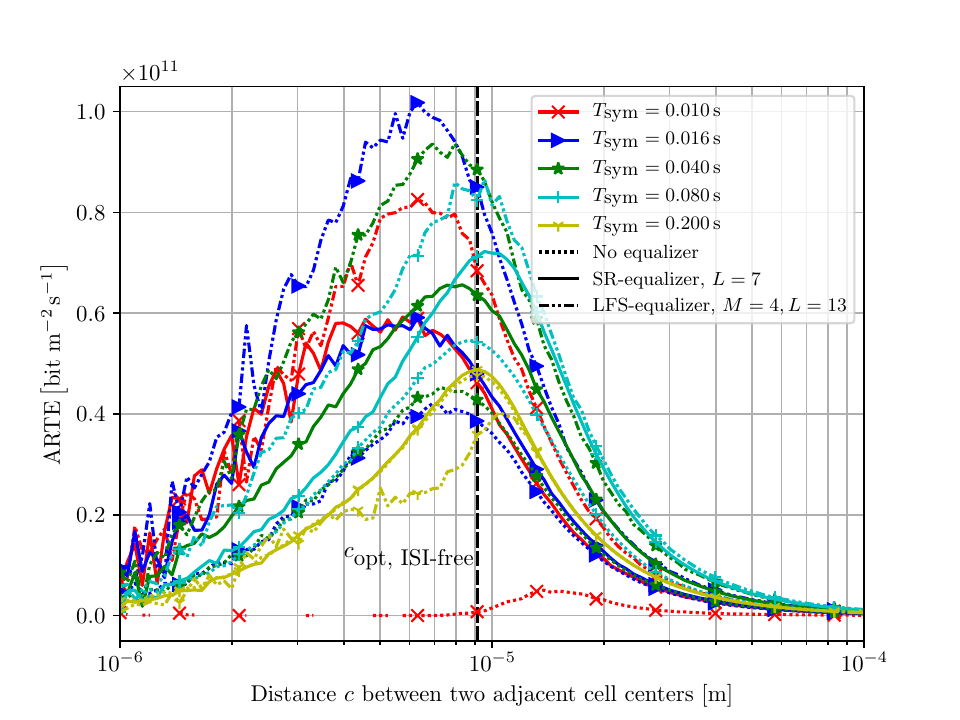}
        \caption{ARTE as a function of cell center distance $\gls{r0minHex}$ for different symbol durations $T_{\mathrm{sym}}$ and different equalization methods. The \ac{ARTE} as defined in \Equation{eq:performance:achievableRate} is obtained by Monte Carlo simulation.}\label{graphic:evaluation:rate_ISI}
    \end{minipage}
\end{figure*}

In this section, we show the \ac{ARTE} as defined in \Equation{eq:performance:achievableRate} for various symbol durations. As small symbol durations lead to \ac{ISI}, we also show results obtained by utilizing appropriate equalizers proposed for MC \cite{cao2020fractionally}, which mitigate the \ac{ISI}.

\scaleSubsubsection
\subsubsection{Design of the Equalizers}\label{section:evaulation:ISI:eq}
\scaleSubsubsectionBelow

We evaluate the \ac{BER} and \ac{ARTE} for two linear equalization schemes, namely a linear symbol rate (SR)-equalizer and a linear fractionally spaced (LFS)-equalizer \cite{cao2020fractionally}\footnote{In \cite{huang2021frequency}, frequency domain equalizers for \ac{MC} systems are proposed. Additionally, the authors of \cite{cao2020fractionally} investigate non-linear fractionally spaced decision feedback (FSDF)-equal\-iz\-ers, whose \ac{BER} performance was shown to be similar to that of LFS-equalizers for the \ac{MC} scenarios considered in \cite{cao2020fractionally}.
To avoid overcrowding \Figures{graphic:evaluation:BER_ISI}{graphic:evaluation:rate_ISI}, we only include results for linear equalization schemes. However, we note that more sophisticated equalizers can yield improved \ac{BER} and \ac{ARTE} performance.}, and compare them to a scheme without equalization.
Here, the equalizers use $L \times M$ samples of the received signal, where $L$ and $M$ denote the number of considered symbols and the samples per symbol, respectively. In particular, $(L-1)/2 \times M$ samples are obtained before \gls{samplingTime}, one sample at \gls{samplingTime}, and $(L+1)/2 \times M - 1$ samples after \gls{samplingTime}. The samples are spaced uniformly by $T_{\mathrm{sym}}/M$. For the SR-equalizer $M=1$ is used by definition. A general mathematical expression for the output signal of the linear equalizers can be found in \cite[Eq. (4)]{cao2020fractionally}.
Here, for each considered value of \gls{r0minHex}, the value of $T_{\mathrm{sym}}$ is chosen such that the \ac{ARTE} is maximized for the two depicted equalization schemes, i.e., an increase and decrease of $T_{\mathrm{sym}}$ to larger and lower values, respectively, results in lower \acp{ARTE}.

\scaleSubsubsection
\subsubsection{Simulation Results considering ILI and ISI}\label{section:evaulation:ISI:results}
\scaleSubsubsectionBelow

\Figures{graphic:evaluation:BER_ISI}{graphic:evaluation:rate_ISI} depict the \ac{BER} and the \ac{ARTE}, respectively, as functions of the hexagonal cell center distance \gls{r0minHex}, the symbol duration $T_{\mathrm{sym}}$, and the equalization scheme. Additionally, the \ac{BER} and $\gls{r0minHex}_{\mathrm{opt, ISI-free}}$ of the ISI-free scenario are shown for reference.\par
\Figure{graphic:evaluation:BER_ISI} shows a \ac{BER} of $0.5$ for small \gls{r0minHex}, which decreases for increasing \gls{r0minHex}, similarly to all previously shown \ac{BER} results. However, above a certain value of \gls{r0minHex}, the \ac{BER} increases again. This behavior is caused by an increase of the \ac{ISI}. In particular, as \gls{r0minHex} increases, the RX volume increases. Therefore, molecules of symbols previously sent by \gls{TX0} are more likely to be observed in the current symbol interval. A further increase in \gls{r0minHex} results in a constant error floor. As for large \gls{r0minHex} the \ac{ILI} approaches zero, the level of the error floor is mainly \ac{ISI} dependent. From \Figure{graphic:evaluation:BER_ISI}, we observe that for increasing symbol duration $T_{\mathrm{sym}}$ the \ac{BER} decreases. This is intuitive as an increase in $T_{\mathrm{sym}}$ decreases the \ac{ISI} and therefore improves the information transmission performance.\par
Next, we compare the performance of the different equalization schemes in \Figure{graphic:evaluation:BER_ISI}. In accordance with \cite{cao2020fractionally}, the LFS-equalizer achieves the best performance, i.e, the lowest \ac{BER}, followed by the SR-equalizer, except for $T_{\mathrm{sym}} = 0.2 \, \si{\second}$.
Furthermore, as expected, the scheme which does not perform equalization, exhibits the largest \ac{BER}, again with the exception of $T_{\mathrm{sym}} = 0.2\, \si{\second}$. In particular, for $T_{\mathrm{sym}} = 0.2\, \si{\second}$, the \acp{BER} obtained with the SR-equalizer and without equalization are similar, while the results obtained with the LFS-equalizer are worse\footnote{The computations of the equalization filter coefficients require the inversion of the covariance matrix of the received signal, which becomes numerically instable if the determinant of the matrix approaches zero. The numerical instability, which is more critical for the LFS-equalizer, is the reason for the worse performance of the LFS-equalizer compared to the SR-equalizer for $T_{\mathrm{sym}} = 0.2\, \si{\second}$.}. From this, we conclude that for $T_{\mathrm{sym}} \geq 0.2 \, \si{\second}$ the \ac{BER} performance degradation caused by \ac{ISI} is small. In contrast, for $T_{\mathrm{sym}} < 0.2 \, \si{\second}$, ISI deteriorates the performance as is evident from the comparison with the ISI-free scenario (solid black line), which assumes very long symbol durations.\par
From \Figure{graphic:evaluation:rate_ISI}, we observe that even in the presence of \ac{ISI} the \ac{ARTE} exhibits a unique maximium \ac{w.r.t.} the cell center distance \gls{r0minHex}, which is achieved for $c_{\mathrm{opt}}$. However, $c_{\mathrm{opt}}$ depends on both the symbol duration $T_{\mathrm{sym}}$ and the equalization scheme. \Figure{graphic:evaluation:rate_ISI} shows that the highest \acp{ARTE} are achieved by the LFS-equalizer, followed by the SR-equalizer, and the scheme without equalization. This is intuitive, as the equalizers mitigate the \ac{ISI} and improve the \ac{BER} (cf. \Figure{graphic:evaluation:BER_ISI}), while the other factors in \Equation{eq:performance:achievableRate}, namely $\gls{spatialRate}$ and $\frac{1}{T_{\mathrm{sym}}}$, are independent of the choice of equalizer. Again, as in \Figure{graphic:evaluation:BER_ISI}, $T_{\mathrm{sym}} = 0.2\, \si{\second}$ is an exception. In particular, for $T_{\mathrm{sym}} = 0.2\, \si{\second}$, for no equalizer and the SR-equalizer, the optimal cell center distance equals the one of the ISI-free scenario, i.e., $c_{\mathrm{opt, }T_{\mathrm{sym}}= 0.2\, \si{\second}} = c_{\mathrm{opt, ISI-free}}$. The LFS-equalizer performs worse due to its inferior \ac{BER} performance. From these observations, we conclude that, for the given parameter values, $T_{\mathrm{sym}}= 0.2\, \si{\second}$ is a sufficiently large symbol duration for which \ac{ISI} can be neglected.\par
Next, we concentrate on the impact of $T_{\mathrm{sym}}$ for the scheme without equalization. \Figure{graphic:evaluation:rate_ISI} shows that the largest \ac{ARTE} is reached for $T_{\mathrm{sym}} = 0.08\, \si{\second}$ at a cell center distance $c_{\mathrm{opt, }T_{\mathrm{sym}}= 0.08\, \si{\second}}$. We observe that decreasing the symbol duration $T_{\mathrm{sym}}$ changes $c_{\mathrm{opt}}$ and decreases the maximum \ac{ARTE}. In particular, for $T_{\mathrm{sym}}=0.04\, \si{\second}$ and $T_{\mathrm{sym}}=0.016\, \si{\second}$ it is optimal \ac{w.r.t.} the overall \ac{ARTE} to employ smaller cell areas compared to $T_{\mathrm{sym}}= 0.08\, \si{\second}$, while for $T_{\mathrm{sym}}=0.01\, \si{\second}$ larger areas are favorable. The observed non-linear dependence of the \ac{ARTE} on $T_{\mathrm{sym}}$ emphasizes the importance of the analysis presented here.
Next, we compare the symbol durations $T_{\mathrm{sym, max}}$, which yield the largest \acp{ARTE} for the considered schemes. We observe that an increase in the number of samples utilized by the equalizer improves its ability to handle interference and therefore can decrease $T_{\mathrm{sym, max}}$ as well as $c_{\mathrm{opt}}$. In particular, $T_{\mathrm{sym, max}}=0.08\, \si{\second}$, $T_{\mathrm{sym, max}}=0.08\, \si{\second}$, and $T_{\mathrm{sym, max}}=0.016\, \si{\second}$ for the scheme without an equalizer, the SR-equalizer, and the LFS-equalizer, respectively. We note that for large cell center distances \gls{r0minHex}, $T_{\mathrm{sym}}=0.08\, \si{\second}$ yields the highest \acp{ARTE} despite the significant \ac{ISI}.\par
We observe that neither the largest nor the smallest symbol durations and cell center distances yield the maximum \ac{ARTE}. Hence, to optimize information transmission in a multi-link \ac{MC} system, i.e., to reach the maximum \ac{ARTE}, both $T_{\mathrm{sym}}$ and \gls{r0minHex} should be judiciously chosen. Finally, we observe from \Figure{graphic:evaluation:rate_ISI} that cell center distance $\gls{r0minHex}_{\mathrm{opt, ISI-free}}$, which maximizes the \ac{ARE}, is close to the optimal cell center distances in the \ac{ISI} impaired scenarios. Hence, by analyzing and optimizing the \ac{ARE}, which is computationally feasible, a cell center distance which incurs a small performance loss in terms of the \ac{ARTE} compared to the optimal cell center distance can be found.

\scaleSection
\section{Conclusion}\label{sec:conclusion}
\scaleSectionBelow
In this paper, we focused on the spatial dimension of \ac{MC} systems. We considered a \ac{3D} system with multiple independent and spatially distributed point-to-point transmission links, where the \ac{TXs} and the \ac{RXs} were positioned according to a regular grid pattern. We proposed the \ac{ARE} and \ac{ARTE} as new performance metrics for \ac{MC} systems to characterize how efficiently given \ac{TX} and \ac{RX} areas are utilized for information transmission. Since space is a valuable resource in multi-link \ac{MC} scenarios, the \ac{ARE} and \ac{ARTE} are relevant performance metrics. To evaluate the considered system in terms of its \ac{ARE}, we developed analytical expressions for the \ac{CIRs} of all existing \ac{TX}-\ac{RX} links, the optimal and suboptimal threshold values for detection, and the \ac{BER} of a single transmission link. Finally, we obtained quantitative results for the optimal spatial link density maximizing the \ac{ARE} of \ac{MC} systems. We showed that the maximum \ac{ARE} is strongly dependent on the number of released molecules and the concentration of the background noise, whereas the considered grid and the approximation error in the calculation of the threshold value have only a minor impact on the maximum \ac{ARE}. Furthermore, we revealed that the optimal link density depends on the number of released molecules and the diffusion coefficient. On the other hand, our obtained results showed that the background noise concentration and the grid structure have a negligible impact on the optimal link density.

In this paper, we have analyzed the \ac{ARE} and \ac{ARTE} for a specific \ac{MC} system under idealized assumptions used such as perfect alignment of \ac{TXs} and \ac{RXs}, perfect time synchronization, uniform flow, and transparent \ac{RXs}. These assumptions can be relaxed in future work to extend the analysis of the \ac{ARE} and \ac{ARTE} to different \ac{MC} channels, unsynchronized transmission, more realistic \ac{RX} models, e.g., absorbing \ac{RXs} \cite{yaylali2023channel}, and different \ac{TX} and \ac{RX} location models including random location models.

\scaleSection
\appendix
\section{}
\scaleSubsection
\subsection{Derivation of the \ac{CIR}}\label{section:appendix:CIRProof}
\scaleSubsectionBelow
The molecule concentration $C(x, x_{\textnormal{TX}_i}, y, y_{\textnormal{TX}_i}, z, z_{\textnormal{TX}}, t, t_{0})$ at position $(x,y,z)$ for the case considered in \Proposition{prop:cir}, i.e., transmitter position
 $\boldsymbol{p}_{\gls{curTrans}} = (x_{\textnormal{TX}_i}, y_{\textnormal{TX}_i}, z_{\textnormal{TX}})$ and release time $t_{0} = 0$, is given in \cite[Eq. (18)]{Jamali2019ChannelMF}. $C(x, x_{\textnormal{TX}_i}, y, y_{\textnormal{TX}_i}, z, z_{\textnormal{TX}_i}, t, t_{0})$ can be transformed to a cylindrical coordinate system by \mbox{$x = r \cos(\varphi)$}, \mbox{$y = r \sin(\varphi)$}, \mbox{$x_{\textnormal{TX}_i} = r_{\gls{curTrans}} \cos(\varphi_{\gls{curTrans}})$}, \mbox{$y_{\textnormal{TX}_i} = r_{\gls{curTrans}} \sin(\varphi_{\gls{curTrans}})$}. Here, $\varphi$ denotes the angle in the $xy$-plane in relation to the positive $x$-axis. We obtain:
  \scaleAlign
  \begin{align}
    &C(r, r_{\gls{curTrans}}, \varphi, \varphi_{\gls{curTrans}}, z, z_{\mathrm{TX}}, t) = \frac{1}{(4 \pi \gls{diffusion} t)^{3/2}} \nonumber \\
    &\quad \times \, \exp\mkern-4.5mu\left(\mkern-4.5mu-\frac{r^2\mkern-2.5mu+\mkern-2.5mur_{\gls{curTrans}}^2 \mkern-2.5mu-\mkern-2.5mu 2 r_{\gls{curTrans}} r \cos(\varphi-\varphi_{\gls{curTrans}}) \mkern-2.5mu+\mkern-2.5mu (z\mkern-2.5mu-\mkern-2.5muz_{\mathrm{TX}}\mkern-2.5mu-\mkern-2.5mu\gls{flow} t)^2}{4 \gls{diffusion} t} \right).
  \end{align}
  Furthermore, since the center of the circular cross-section of \gls{RX0} is located at $r=0$, the expected number of counted molecules is obtained as
  \scaleAlign
  \begin{align}
    &\textnormal{CIR}_{\gls{curTrans}}(t)= \int \limits_0^{\gls{SRX}} \int \limits_0^{2 \pi} \int \limits_{z =z_{\mathrm{S}}}^{z_{\mathrm{E}}}  C(r, r_{\gls{curTrans}}, \varphi, \varphi_{\gls{curTrans}}, z, z_{\mathrm{TX}}, t) r \, \mathrm{d} z  \,\mathrm{d}  \varphi   \,\mathrm{d}  r \nonumber \\[-0.2cm]
    &\overset{(a)}{=} \frac{1}{4 \gls{diffusion} t} \left(\erf\left(m_0\right)- \erf\left(m_1\right)\right) \nonumber \\
    &\qquad \times \,\int \limits_0^{\gls{SRX}}  \mathrm{I}_0\left(\frac{r_{\gls{curTrans}} r}{2 \gls{diffusion} t}\right) \exp\left(-\frac{r^2+r_{\gls{curTrans}}^2}{4 \gls{diffusion} t} \right) r \,\mathrm{d}  r \nonumber  \\[-0.1cm]
    &\overset{(b)}{=} \frac{1}{4 \gls{diffusion} t} \left(\erf\left(m_0\right)- \erf\left(m_1\right)\right) \exp\left(-\frac{r_{\gls{curTrans}}^2}{4 \gls{diffusion} t} \right)  \nonumber \\
    &\qquad \times \,\mkern-5mu\int\limits_0^{\gls{SRX}} \, \sum_{k=0}^{k_{\mathrm{max}} = \infty} \exp\left(-\frac{r^2}{4 \gls{diffusion} t} \right) \frac{\left(\frac{r_{\gls{curTrans}} r}{4 \gls{diffusion} t}\right)^{2 k} }{k! \Gamma(k+1)} r \,\mathrm{d}  r  \nonumber \\[-0.1cm]
    &\overset{(c)}{=} \frac{1}{4 \gls{diffusion} t} \left(\erf\left(m_0\right)- \erf\left(m_1\right)\right) \exp\left(-\frac{r_{\gls{curTrans}}^2}{4 \gls{diffusion} t} \right) \mkern-5mu \nonumber \\
    &\qquad \times \,\int \limits_0^{\gls{SRX}} \, \sum_{k=0}^{k_{\mathrm{max}} = \infty}\mkern-15mu \exp\left(-\frac{r^2}{4 \gls{diffusion} t} \right) \frac{\left(\frac{r_{\gls{curTrans}} r}{4 \gls{diffusion} t}\right)^{2 k} }{(k!)^2} r \,\mathrm{d}  r \;,
    \label{eq:math_section:CIR}
  \end{align}
  where we substituted $m_0 = \frac{z_{\mathrm{TX}} + \gls{flow} t - z_{\mathrm{S}}}{\sqrt{4 \gls{diffusion} t}}$ and $m_1 = \frac{z_{\mathrm{TX}} +\gls{flow} t - z_{\mathrm{E}}}{\sqrt{4 \gls{diffusion} t}}$ and exploited $(a)$ $\erf(x) = \frac{2}{\sqrt{\pi}} \int_0^x \exp\left(-y^2 \right)  \,\mathrm{d}  y$ and $  \mathrm{I}_0(x) = \frac{1}{\pi} \int_0^{\pi} \exp\left(x \cos(\varphi)\right) \,\mathrm{d}  \varphi,$ $(b)$ the series expansion of $ \mathrm{I}_0(x)$ \cite[Eq. (9.6.10)]{abramowitz1964handbook}, $(c)$ the relation between the Gamma function and the factorial $\Gamma(n) = (n-1)!$. Finally, we exploit $\gamma(a,x) = \int_0^x y^{a-1} \exp(-y) \,\mathrm{d} y$ to obtain \Equation{eq:math_ir_general}. Here, $\mathrm{I}_0(x)$ denotes the zeroth order modified Bessel function of the first kind.

\scaleSubsection
  \subsection{Proof of the Mismatch Error Bound of the CIR for Finite $k_{\mathrm{max}}$ }\label{section:appendix:kmax_truncation}
  \scaleSubsectionBelow
    The bound of the mismatch error $\textnormal{CIR}_{\gls{curTrans}}(t = \gls{samplingTime}) \given{_{k_{\mathrm{max}} = \infty}} \, - \, \textnormal{CIR}_{\gls{curTrans}}(t = \gls{samplingTime}) \given{_{k_{\mathrm{max}} = k'}}$ is obtained as
  \scaleAlign
  \begin{align}
    &\textnormal{CIR}_{\gls{curTrans}}(t = \gls{samplingTime}) \given{_{k_{\mathrm{max}} = \infty}} \quad - \quad \textnormal{CIR}_{\gls{curTrans}}(t = \gls{samplingTime}) \given{_{k_{\mathrm{max}} = k'}} \nonumber \\[-0.02cm]
    &= \frac{1}{2} \mkern-4.5mu\left(\mkern-4.5mu\erf\mkern-4.5mu\left(\mkern-4.5mu\frac{z_{\textnormal{TX}} + \gls{flow}\gls{samplingTime} - z_{\mathrm{S}}}{\sqrt{4 D\gls{samplingTime}}}\mkern-4.5mu\right)\mkern-4.5mu- \erf\mkern-4.5mu\left(\mkern-4.5mu\frac{z_{\textnormal{TX}} +\gls{flow}\gls{samplingTime} - z_{\mathrm{E}}}{\sqrt{4 \gls{diffusion}\gls{samplingTime}}}\mkern-4.5mu\right)\mkern-2.5mu\mkern-2.5mu\right)\mkern-4.5mu\nonumber \\
    &\quad \times \,\exp\mkern-4.5mu\left(\mkern-4.5mu-\frac{r_{\gls{curTrans}}^2}{4 \gls{diffusion}\gls{samplingTime}}\mkern-2.5mu\right)\mkern-4.5mu \sum_{k=k'+1}^{\infty} \frac{\mkern-4.5mu\left(\mkern-4.5mu\frac{r_{\gls{curTrans}}^2}{4 \gls{diffusion}\gls{samplingTime}}\mkern-2.5mu\right)\mkern-4.5mu^k}{(k!)^2} \gamma\mkern-4.5mu\left(\mkern-4.5muk+1,\frac{\gls{SRX2}}{4 \gls{diffusion}\gls{samplingTime}}\mkern-2.5mu\right)\mkern-4.5mu \nonumber \\
    &\overset{(a)}{=} w_0 w_1 \sum_{k=k'+1}^{\infty} \frac{w_2^k}{(k!)^2} \frac{\gamma\left(k+1,w_3\right)}{\Gamma\left(k+1\right)} \Gamma\left(k+1\right) \nonumber \\
    &\overset{(b)}{\leq} w_0 w_1 \sum_{k=k'+1}^{\infty} \frac{w_2^k}{(k!)} \left(1-\exp\left(-w_3\right)\right)^{k+1} \nonumber \\
    &\overset{(c)}{=} w_0 w_1 (1-\exp(-w_3)) \sum_{k=k'+1}^{\infty} \frac{\left( w_2 - w_2 \exp(-w_3) \right)^k}{k!} \nonumber\\
    &\overset{(d)}{=} w_5 \left( \sum_{k=0}^{\infty} \frac{w_4^k}{k!}  - \sum_{k=0}^{k'} \frac{w_4^k}{k!} \right) \nonumber \\
    &\overset{(e)}{=} \mkern-2.5mu w_5 \mkern-4.5mu\left(\mkern-4.5mu \exp(w_4) - \exp(w_4) \frac{\Gamma( k' + 1 ,w_4)}{\Gamma(k'+1)} \mkern-4.5mu\right)\mkern-4.5mu  \mkern-2.5mu \nonumber \\
    &\overset{(f)}{=}\mkern-2.5mu \frac{w_5 \exp(w_4)}{k'!} \gamma(k' + 1, w_4) \mkern-2.5mu\overset{!}{<}\mkern-2.5mu \eta \; \textnormal{CIR}_{0}(t = \gls{samplingTime}), \label{inequ:error_k_max}
  \end{align}
  where we substitute in $(a)$ $w_0 = \frac{1}{2} \left(\erf\left(\frac{z_{\textnormal{TX}} + \gls{flow}\gls{samplingTime} - z_{\mathrm{S}}}{\sqrt{4 D\gls{samplingTime}}}\right)- \erf\left(\frac{z_{\textnormal{TX}} +\gls{flow}\gls{samplingTime} - z_{\mathrm{E}}}{\sqrt{4 \gls{diffusion}\gls{samplingTime}}}\right)\right)$, $w_1 = \exp\left(-\frac{r_{\gls{curTrans}}^2}{4 \gls{diffusion}\gls{samplingTime}}\right)$, $w_2 = \frac{r_{\gls{curTrans}}^2}{4 \gls{diffusion}\gls{samplingTime}}$, and $w_3 = \frac{\gls{SRX2}}{4 \gls{diffusion}\gls{samplingTime}}$, and apply in $(b)$ $\Gamma\left(k\right) = (k-1)!$ and Lemma~\ref{Lem:upper_bound_gamma}. Subsequently, after applying the exponential rule $x^{m+1} = x^m x$ in $(c)$, we substitute in $(d)$ $w_4 = w_2 - w_2 \exp(-w_3)$ and $w_5 = w_0 w_1 (1-\exp(-w_3))$ and split the sum. Furthermore, we exploit in $(e)$ the power series definition of the exponential function $\exp(x) = \sum_{k=0}^{\infty} \frac{x^k}{k!}$ and the \ac{CDF} of a Poisson distribution $\sum_{k = 0}^{x-1} \frac{{\lambda}^{k} \exp\left(-\lambda\right)}{k!} = \frac{\Gamma( x ,\lambda)}{\Gamma(x )}, \; \mathrm{with}\; x \in \mathbb{N}$ and rate $\lambda > 0$ . Here, $\Gamma(a,b)$ denotes the upper incomplete Gamma function. Finally, we apply in $(f)$ $\Gamma\left(k\right) = (k-1)!$ and $\Gamma\left(a\right) = \Gamma(a,b) + \gamma(a,b)$, which holds by definition. This concludes the proof.

  \begin{lem}\label{Lem:upper_bound_gamma}
    Given $z > 1$ and $x > 0$, $\frac{\gamma\left(z,x\right)}{\Gamma\left(z\right)}$ is upper bounded as follows
    \scaleAlign
    \begin{align}
       \frac{\gamma\left(z,x\right)}{\Gamma\left(z\right)} \leq \left(1-\exp\left(-x\right)\right)^{z} \;.
       \label{lemma1_equation}
    \end{align}
  \end{lem}

\scaleSubsection
\subsection{Proof of Lemma~\ref{Lem:upper_bound_gamma}}\label{proof_lemma}
\scaleSubsectionBelow
Rearranging \Equation{lemma1_equation} yields $\gamma\left(z,x\right) \left(1-\exp\left(-x\right)\right)^{z} \leq \Gamma\left(z\right)$, for which equality holds for $x \rightarrow \infty$. Therefore, to prove Lemma~\ref{Lem:upper_bound_gamma}, it is sufficient to show that $f_z(x) = \gamma\left(z,x\right) \left(1-\exp\left(-x\right)\right)^{z}$ is monotonically increasing in $x$, given $x>0$ and $z>1$, i.e.,
\begin{align}
  &\frac{\partial }{\partial x} f_z(x) = \exp\left(x z\right) \left(\exp(x) -1\right)^{-z-1}\nonumber \\
  &\quad \times \, \left( x^{z-1} (1-\exp(-x)) - z \int\limits_0^xy^{z-1} \exp(-y) \ensuremath{\mathrm{d}y} \right) > 0 . \label{}
\end{align}
Since $\exp\left(x z\right) \left(\exp(x) -1\right)^{-z-1} > 0$, we need to show that $g_z(x) = x^{z-1} (1-\exp(-x)) - z \int_0^x y^{z-1} \exp(-y) \ensuremath{\mathrm{d}y} > 0$, which is achieved by showing that $g_z(0) = 0$ and $\frac{\partial }{\partial x} g_z(x) = (z-1) x^{z-2} \exp(-x) (\exp(x)-1-x) > 0$ for $x>0$. In fact, $(z-1) x^{z-2} \exp(-x) > 0$ for $z>1$ and $x>0$. Furthermore, $ \exp(x) > 1+x$ holds as $\exp(x) = 1 + x + \frac{x^2}{2} + ... $. This concludes the proof.

\scaleSubsection
\subsection{Proof of Existence of Optimal Threshold Value}\label{section:appendix:ThresholdProof}
\scaleSubsectionBelow
The existence of a unique threshold level $\gls{threshold}_{\mathrm{opt}, \gls{nrTrans}-1}$ can be proven by showing that the fraction in \Equation{eq:math_section_ML} is monotonically increasing in $\gls{r}$, i.e., for low values of \gls{r}, i.e., $\gls{r} <  \gls{threshold}_{\mathrm{opt}, \gls{nrTrans}-1}$, the fraction is always smaller than $1$, and for large values of \gls{r}, i.e., $\gls{r} \geq  \gls{threshold}_{\mathrm{opt}, \gls{nrTrans}-1}$, the fraction is always at least equal to $1$. This can be proven using the same steps as in a similar proof in \cite[Appendix]{Jamali2018NonCoherentDF}.
\begin{lem}\label{Lem:func}
If function $\frac{f_l(t)}{g_p(t)}$ is monotonically increasing, then function $\frac{\sum_l f_l(t)}{\sum_p g_p(t)}$ is also monotonically increasing.
\end{lem}
\begin{IEEEproof}
Please refer to the proof in \cite[Appendix]{Jamali2018NonCoherentDF}.
\end{IEEEproof}
For Lemma~\ref{Lem:func}, we have to show that
\scaleAlign
\begin{align}
\frac{f_l(t)}{g_p(t)} &= \frac{{(\gls{csmean} + \gls{siui}_{,l}\gls{cmeaniuiTrans} + \gls{cmeann})}^{\gls{r}} \exp\left(-(\gls{csmean} + \gls{siui}_{,l}\gls{cmeaniuiTrans} + \gls{cmeann})\right)}{{(\gls{siui}_{,p}\gls{cmeaniuiTrans} + \gls{cmeann})}^{\gls{r}} \exp\left(-(\gls{siui}_{,p}\gls{cmeaniuiTrans} + \gls{cmeann})\right)} \nonumber \\
&= \left(\frac{{\gls{csmean} + \gls{siui}_{,l}\gls{cmeaniuiTrans} + \gls{cmeann}} }{{\gls{siui}_{,p}\gls{cmeaniuiTrans} + \gls{cmeann}} }\right)^{\gls{r}} \nonumber \\
&\qquad \times \, \exp\left(-(\gls{csmean})\right)  \exp\left(\gls{siui}_{,p}\gls{cmeaniuiTrans} - \gls{siui}_{,l}\gls{cmeaniuiTrans}\right)
\end{align}
is monotonically increasing in $\gls{r}$ for all combinations of $l$ and $p$, where $\gls{siui}_{,l}$ and $\gls{siui}_{,p}$ denote two possible realizations of $\gls{siui}$. Therefore, as long as
\scaleAlign
\begin{align}
  \frac{{\gls{csmean} +  \overbrace{\gls{siui}_{,l}\gls{cmeaniuiTrans}}^{\geq 0} + \gls{cmeann} }}{{\gls{siui}_{,p}\gls{cmeaniuiTrans} +\gls{cmeann} } } \geq \frac{{\gls{csmean}} }{{\gls{siui}_{,p}\gls{cmeaniuiTrans} } }  \nonumber \\ \geq \frac{\gls{csmean}}{\text{max}\{\gls{siui}\}\gls{cmeaniuiTrans}}
  = \frac{\gls{csmean}}{\mathbf{1}\gls{cmeaniuiTrans}} = \textrm{SINR}_{\textrm{worst}} > 1
  \label{eq:app:sinr_condition}
\end{align}
 is valid, the existence of a unique threshold value can be guaranteed, where $\mathbf{1}$ denotes the all-ones row vector. Note that $\textrm{SINR}_{\textrm{worst}} > 1$ is a sufficient, but not a necessary condition, i.e., for $\textrm{SINR}_{\textrm{worst}} \leq 1$, a single threshold value might still be equivalent to the \ac{ML} decision rule given in \Equation{eq:math_section_ML}. However, this can not be guaranteed%
 {\footnote{A similar proof was given in the Appendix of \cite{Jamali2018NonCoherentDF}. Contrary to the statement there, $\frac{f_n(r[k])}{g_m(r[k])}$ is not always a monotonically increasing function in $r[k]$. However, imposing the mild condition $\frac{\textnormal{min}(\overline{c}_{\textnormal{s}})}{\textnormal{max}(\overline{c}_{\textnormal{n}})} > 1$ (i.e., the worst-case \ac{SNR} is larger than $0\,\mathrm{dB}$) guarantees the monotonicity of $\frac{f_n(r[k])}{g_m(r[k])}$ in $r[k]$ and the rest of the proof remains valid.}}.

\scaleSubsection
\subsection{Derivation of the Bit Error Rate}\label{section:appendix:BERProof}
\scaleSubsectionBelow
    The \ac{BER} $\gls{errorProb}$ in the presence of \ac{ILI}, i.e., $\gls{nrTrans} > 1$ can be derived as follows:
  \scaleAlign
    \begin{align}
      \gls{errorProb}&\overset{(a)}{=} \frac{1}{2^{\gls{nrTrans}-1}}  \sum\limits_{\gls{siui} \in \mathcal{M}} \left( \frac{1}{2} \prob{ \gls{r} < \gls{threshold}' \given \gls{siui}, \gls{ss} = 1} \right.\nonumber\\
      &\qquad \left. + \frac{1}{2} \prob{\gls{r} \geq \gls{threshold}' \given \gls{siui}, \gls{ss} = 0}\right) \nonumber \\
      &\overset{(b)}{=} \frac{1}{2^{\gls{nrTrans}-1}}  \sum\limits_{\gls{siui} \in \mathcal{M}} \left( \frac{1}{2}  \sum_{\gls{r} = 0}^{\gls{threshold}'-1} \fpdf[\gls{r}]{\gls{r} \given \gls{siui}, \gls{ss} = 1} \right.\nonumber\\
      &\qquad \left. + \frac{1}{2} \sum_{\gls{r} = \gls{threshold}'}^{\infty} \fpdf[\gls{r}]{\gls{r} \given \gls{siui}, \gls{ss} = 0}\right) \nonumber \\
      &\overset{(c)}{=} \frac{1}{2^{\gls{nrTrans}-1}}  \sum\limits_{\gls{siui} \in \mathcal{M}} \nonumber \\
      &\left(\mkern-4.5mu \frac{1}{2} \mkern-4.5mu \sum_{\gls{r} = 0}^{\gls{threshold}'-1}  \poisDist{\gls{csmean}\mkern-2.5mu + \mkern-2.5mu\gls{siui}\gls{cmeaniuiTrans} + \gls{cmeann}}{\gls{r}}  \right. \nonumber \\
      & \left.  + \frac{1}{2} \biggl(1 - \sum_{\gls{r} = 0}^{\gls{threshold}'-1} \poisDist{\gls{siui}\gls{cmeaniuiTrans} + \gls{cmeann}} {\gls{r}}\mkern-5.5mu\biggr)\mkern-5.5mu\right)\mkern-4.5mu ,
    \end{align}
    where we exploit in $(a)$ the threshold detection rule \Equation{eq:math_section:threshold}, in $(b)$ the fact that \gls{r} is an integer value, and in $(c)$ the mass function property $\sum_{\gls{r}} \fpdf[\gls{r}]{\gls{r}} = 1$ and the fact that \gls{r} is Poisson distributed.  Finally, we exploit $\sum_{k = 0}^{x-1} \poisDist{\lambda}{k} = \frac{\Gamma( x ,\lambda)}{\Gamma(x )} = \mathcal{Q}(x ,\lambda), \; \mathrm{with}\; x>0$ to obtain \Equation{eq:performance:error_prob}. Here, $\frac{\Gamma(a,b)}{\Gamma(a)} = \mathcal{Q}(a,b)$ denotes the regularized Gamma function.
    In the absence of \ac{ILI}, i.e., $\gls{nrTrans} = 1$, \Equation{eq:performance:error_prob} simplifies to \Equation{eq:performance:error_prob_IUI_free}.

\bibliographystyle{IEEEtran}
\bibliography{literature}
\end{document}